\documentclass{bmcart}
\usepackage[utf8]{inputenc}
\usepackage{threeparttable}
\usepackage{multirow}
\usepackage{array}
\usepackage[pdftex]{graphicx}
\usepackage{amsthm,amsmath}
\usepackage{threeparttable}
\usepackage{multirow}
\usepackage{multicol}
\usepackage{float}
\usepackage{color}
\usepackage{hyperref}

\startlocaldefs
\endlocaldefs

\begin{document}
\begin{frontmatter}
\begin{fmbox}
\dochead{Research}
\title{Latent Class Model with Application to Speaker Diarization}
    \author[
    addressref={aff1},
    corref={aff1},
    email={heliang@mail.tsinghua.edu.cn}
    ]{\inits{LH}\fnm{Liang} \snm{He}}
    \author[
    addressref={aff1},
    email={chenxianhong@mail.tsinghua.edu.cn}
    ]{\inits{XC}\fnm{Xianhong} \snm{Chen}}
    \author[
    addressref={aff1},
    email={xucan@mail.tsinghua.edu.cn}
    ]{\inits{CX}\fnm{Can} \snm{Xu}}
    \author[
    addressref={aff1},
    email={liu-yi15@mails.tsinghua.edu.cn}
    ]{\inits{JY}\fnm{Yi} \snm{Liu}}
    \author[
    addressref={aff1},
    email={liuj@mail.tsinghua.edu.cn}
    ]{\inits{JL}\fnm{Jia} \snm{Liu}}
    \author[
    addressref={aff2},
    email={mike.johnson@uky.edu}
    ]{\inits{MTJ}\fnm{Michael~T} \snm{Johnson}}

    \address[id=aff1]{
      \orgname{Department of Electronic Engineering, Tsinghua University},
      \street{Zhongguancun Street},
      \postcode{100084},
      \city{Beijing},
      \cny{China}
    }
    \address[id=aff2]{
      \orgname{Electrical and Computer Engineering, University of Kentucky},
      \street{453E FPAT},
      \postcode{40506-0046},
      \city{Lexington, Kentucky},
      \cny{USA}
    }
\end{fmbox}

\begin{abstractbox}
\begin{abstract}
In this paper, we apply a latent class model (LCM) to the task of speaker diarization.
LCM is similar to Patrick Kenny's variational Bayes (VB) method in that it uses soft information and avoids premature hard decisions in its iterations.
In contrast to the VB method, which is based on a generative model,
LCM provides a framework allowing both generative and discriminative models.
The discriminative property is realized through the use of i-vector (Ivec), probabilistic linear discriminative analysis (PLDA), and a support vector machine (SVM) in this work.
Systems denoted as LCM-Ivec-PLDA, LCM-Ivec-SVM, and LCM-Ivec-Hybrid are introduced.
In addition, three further improvements are applied to enhance its performance.
1) Adding neighbor windows to extract more speaker information for each short segment.
2) Using a hidden Markov model to avoid frequent speaker change points.
3) Using an agglomerative hierarchical cluster to do initialization and present hard and soft priors, in order to overcome the problem of initial sensitivity.
Experiments on the National Institute of Standards and Technology Rich Transcription 2009 speaker diarization database, under the condition of a single distant microphone, show that the diarization error rate (DER) of the proposed methods has substantial relative improvements compared with mainstream systems.
Compared to the VB method, the relative improvements of LCM-Ivec-PLDA, LCM-Ivec-SVM, and LCM-Ivec-Hybrid systems are 23.5$\%$, 27.1$\%$, and 43.0$\%$, respectively.
Experiments on our collected database, CALLHOME97, CALLHOME00 and SRE08 short2-summed trial conditions also show that the proposed LCM-Ivec-Hybrid system has the best overall performance.
\end{abstract}


\end{abstractbox}

\end{frontmatter}

\section{Introduction}
\renewcommand{\arraystretch}{1.2}

Speaker diarization task aims to address the problem of "who spoke when" in an audio stream by splitting the audio into homogeneous regions labeled with speaker identities \cite{Miro2012}.
It has a wide application in automatic audio indexing, document retrieving and speaker-dependent automatic speech recognition.

In the field of speaker diarization, variational Bayes (VB) proposed by Patrick Kenny \cite{Kenny2010,Kenny2008,Reynolds2009,Valente2005} and VB-hidden Markov model (HMM) introduced by Mireia Diez \cite{Diez2018} have become the state-of-the-art approaches.
This system has two characteristics.
First, unlike mainstream approaches (i.e. segmentation and clustering approaches, discussed in the following section), it uses a fixed length segmentation instead of speaker change point detection to do speaker segmentation, dividing an audio recording into uniform and short segments.
These segments are short enough that they can be regarded as each containing only one speaker.
This type of segmentation leaves the difficulty to the clustering stage and requires a better clustering algorithm that includes temporal correlation.
Second, the VB approach utilizes a soft clustering approach that avoids premature hard decisions.
Despite its accuracy, there are still some deficiencies of the approach.
The VB approach is a single-objective method.
Its goal is to increase the overall likelihood, which is based on a generative model, not to distinguish speakers.
Furthermore, because the segmented segments are very short, the probability that an individual segment occurs given a particular speaker
is inaccurate and may degrade system performance.
In addition, some researchers have also noted that the VB system is very sensitive to its initialization conditions \cite{Bulut2015}.
For example, if one speaker dominates the recording, a random prior tends to result in assigning the segments to each speaker evenly, leading to a poor result.

In this paper, to address the drawbacks of VB, we apply a latent class model (LCM) to speaker diarization.
LCM was initially introduced by Lazarsfeld and Henry \cite{LCM}.
It is usually used as a way of formulating latent attitudinal variables from dichotomous survey items \cite{LCMclustering2002,LCMusage}.
This model allows us to compute $p(\mathcal{X}_m, \mathcal{Y}_s, i_{ms})$,
which represents the likelihood that both the segment representation $\mathcal{X}_m$ and the estimated class representation $\mathcal{Y}_s$ are from the same speaker, in a more flexible and discriminative way.
We introduce the probabilistic linear discriminative analysis (PLDA) and support vector machine (SVM) into the computation, and propose LCM-Ivec-PLDA, LCM-Ivec-SVM, and LCM-Ivec-Hybrid systems.
Furthermore, to address the problem caused by the shortness of each segment, in consideration of speaker temporal relevance,
we take $\mathcal{X}_m$'s neighbors into account at the data and score levels to improve the accuracy of $p(\mathcal{X}_m,\mathcal{Y}_s)$.
A Hidden Markov model (HMM) is applied to smooth frequent speaker changes.
When the speakers are imbalanced, we use an agglomerative hierarchical cluster (AHC) approach \cite{Han2007} to address the system sensitivity to initialization.

The parameter selection experiments are mainly carried out on the NIST RT09 SPKD database \cite{RT09Eval} and our collected speaker imbalanced database.
In practice, the number of speakers in a meeting or telephone call is relatively easy to be obtained. We assume that this number is known in advance.
RT09 has two evaluation conditions: single distant microphone (SDM), where only one microphone channel is involved;
and multiple distant microphone (MDM), where multiple microphone channels are involved.
In this paper, we mainly consider the speaker diarization task under the SDM condition.
We also conduct performance comparison experiments on the RT09, CALLHOME97 \cite{ldc1997}, CALLHOME00 (a subtask of NIST SRE00) and SRE08 short2-summed trial condition.
Experiment results show that the proposed method has better performance compared with the mainstream systems.

The remainder of this paper is organized as follows.
Section \ref{sec:approaches} describes mainstream approaches and algorithms.
Section \ref{sec:lcm} introduces the latent class model (LCM) and section \ref{sec:realization} realizes the LCM-Ivec-PLDA, LCM-Ivec-SVM, and LCM-Ivec-Hybrid systems.
Further improvements are presented in Section \ref{sec:improvements}.
Section \ref{sec:related} discusses the difference between our proposed methods and related works.
Experiments are carried out and the results are analyzed in Section \ref{sec:exp}. Conclusions are drawn in Section \ref{sec:con}.

\section{Mainstream Approaches and Algorithms} \label{sec:approaches}
Speaker diarization is defined as the task of labeling speech with the corresponding speaker.
The most common approach consists of speaker segmentation and clustering \cite{Miro2012,Zelenak2012}.

The mainstream approach to speaker segmentation is finding speaker change points based on a similarity metric.
This includes Bayesian information criterion (BIC) \cite{BIC2008}, Kullback-Leibler \cite{KL2008}, generalized likelihood ratio (GLR) \cite{GLR2010} and i-vector/PLDA \cite{ivector2015}.
More recently, there are also some metrics based on deep neural networks (DNN) \cite{AANN2009,SCD_DNN2015}, convolutional neural networks (CNN) \cite{SCD_CNN2017,CNN2017later}, and recurrent neural networks (RNN) \cite{RNN2017, LSTM2017}.
However, the DNN related methods need a large amount of labeled data and might suffer from a lack of robustness when working in different acoustic environments.

In speaker clustering, the segments belonging to the same speaker are grouped into a cluster.
The problem of measuring segment similarity remains the same as for speaker segmentation and the metrics described above can also be used for clustering.
Cluster strategies based on hard decisions include agglomerative hierarchical clustering (AHC) \cite{Han2007} and division hierarchical clustering (DHC) \cite{EURECOM2010}.
A soft decision based strategy is the variational Bayes (VB) \cite{Valente2005}, which is combined with eigenvoice modeling \cite{Kenny2010}.
Taking temporal dependency into account, HMM \cite{Diez2018} and hidden distortion models (HDM) \cite{Lapidot2012,Lapidot2016} are successfully applied in speaker diarization.
There are also some DNN based clustering strategies.
In \cite{SC_DNN2016}, a clustering algorithm is introduced by training a speaker separation DNN and adapting the last layer to specific segments.
Another paper \cite{DNNHMMcluster2017} introduces a DNN-HMM based clustering method, which uses a discriminative model rather than a generative model, i.e. replacing GMMs with DNNs, for the estimation of emission probability, achieving better performance.

Some diarization systems based on i-vector, VB or DNN are trained in advance, rely on the knowledge of application scenarios, and require large amount of matched training data. They perform well in fixed conditions.
While some other diarization systems, such as BIC, HMM or HDM, have little prior training. They are condition independent and more robust to the change of conditions.
They perform better if the conditions, such as channels, noises, or languages, vary frequently.


\subsection{Bottom-Up Approach}
The bottom-up approach is the most popular one in speaker diarization \cite{Han2007}, which is often referred to as an agglomerative hierarchical clustering (AHC).
This approach treats each segment, divided by speaker change points, as an individual cluster, and merges a pair of clusters into a new one based on the nearest neighbor criteria.
This merging process is repeated until a stopping criterion is satisfied.
To merge clusters, a similarity function is needed.
When clusters are represented by a single Gaussian or sometimes Gaussian Mixture model (GMM), Bayesian information criterion (BIC) \cite{Zhu2005,IIR2010,ICSI2012} is often adopted.
When clusters are represented by i-vectors, cosine distance \cite{Shum2011} or probabilistic linear discriminant analysis (PLDA) \cite{Sell2014,Lan2016,PldaAhc2014,Zhu2016} is usually used.
The stopping criteria can be based on thresholds, or on a pre-assumed number of speakers, alternatively \cite{Shum2013,Sell2016}.

Bottom-up approach is more sensitive to nuisance variations (compared with the top-down approach), such as speech channel, speech content, or noise \cite{CompareBottomUpTopDown2012}.
A similarity function, which is robust to these nuisance variations, is crucial to this approach.

\subsection{Top-Down Approach}
The top-down approach is usually referred to as a divisive hierarchical clustering (DHC) \cite{EURECOM2010}.
In contrast with the bottom-up approach, the top-down approach first treats all segments as unlabeled.
Based on a selection criterion, some segments are chosen from these unlabeled segments.
The selected segments are attributed to a new cluster and labeled.
This selection procedure is repeated until no more unlabeled segments are left or until the stopping criteria, similar to those employed in the bottom-up approach, is reached.
The top-down approach is reported to give worse performance on the NIST RT database \cite{EURECOM2010} and has thus received less attention.
However, paper \cite{CompareBottomUpTopDown2012} makes a thorough comparative study of these two approaches and demonstrates that these two approaches have similar performance.

The top-down approach is characterized by its high computational efficiency but is less discriminative than the bottom-up approach.
In addition, top-down is not as sensitive to nuisance variation, and can be improved through cluster purification \cite{EURECOM2010}.

Both approaches have common pitfalls. They make premature hard decisions which may cause error propagation.
Although these errors can be fixed by Viterbi resegmentation in next iterations \cite{CompareBottomUpTopDown2012} \cite{Nwe2012}, a soft decision is still more desirable.

\subsection{Hidden Distortion Model}
Different from AHC or DHC, HMM takes temporal dependencies between samples into account.
Hidden distortion model (HDM) \cite{Lapidot2012,Lapidot2016} can be seen as a generalization of HMM to overcome its limitations.
HMM is based on the probabilistic paradigm while HDM is based on the distortion theory.
In HMM, there is no regularization option to adjust the transition probabilities.
In HDM, a regularization of transition cost matrix, used as a replacement of transition probability matrix, is a natural part of the model.
Both HMM and HDM do not suffer from error propagation.
They do re-segmentation via a Viterbi or forward-backward algorithm.
And each iteration may fix errors in previous loops.

\subsection{Variational Bayes}

Variational Bayes (VB) is a soft speaker clustering method introduced to address speaker diarization task \cite{Kenny2010,Valente2005,Diez2018}.
Suppose a recording is uniformly segmented into fixed length segments $\mathcal{X}= \{\mathcal{X}_1,\cdots,\mathcal{X}_m,\cdots,\mathcal{X}_M\}$, where the subscript $m$ is the time index, $1 \leq m \leq M $. $M$ is the segment duration.
Let $\mathcal{Y}=\{\mathcal{Y}_1, \cdots, \mathcal{Y}_s, \cdots, \mathcal{Y}_S\}$ be the speaker representation, where $s$ is the speaker index, $1\leq s \leq S$. $S$ is the speaker number.
$I=\{i_{ms}\}$, where $i_{ms}$ represents whether a segment $m$ belongs to a speaker $s$ or not.
In speaker diarization, $\mathcal{X}$ is the observable data, $\mathcal{Y}$ and $I$ are the hidden variables.
The goal is to find proper $\mathcal{Y}$ and $I$ to maximize $\log p(\mathcal{X})$.
According to the Kullback-Leibler divergence, the lower bound of the log likelihood $\log p(\mathcal{X})$ can be expressed as

\[
\log p(\mathcal{X})\geq \int p(\mathcal{Y},I)\ln \frac{p(X,\mathcal{Y},I)}{p(\mathcal{Y},I)} d(\mathcal{Y},I)
\]

The equality holds if and only if $p(\mathcal{Y},I)=p(\mathcal{Y},I|\mathcal{X})$.
The VB assumes a factorization $p(\mathcal{Y},I) = p(\mathcal{Y}) p(I)$ to approximate the true posterior $p(\mathcal{Y},I|\mathcal{X})$ \cite{Kenny2010}.
Then, $p(\mathcal{Y})$ and $ p(I)$ are iteratively refined to increase the lower bound of $\log p(\mathcal{X})$.
The final speaker diarization label can be assigned according to segment posteriors \cite{Kenny2010}.
The implementation of VB approach is shown in Algorithm 1.
Compared with the bottom-up or top-down approach, the VB approach uses a soft decision strategy and avoids a premature hard decision.

\begin{table}[!ht]
\begin{center}
\begin{tabular}{p{0.9\columnwidth}}
\hline
Algorithm 1: Variational Bayes \\
\hline
1: Voice activity detection and feature extraction \\
2: Speaker segmentation \\
\quad 2.1: Split an audio into $M$ short fixed length segments.   \\
3: Clustering \\
\quad 3.1: \begin{minipage}[t]{0.9\linewidth} For each speaker $s$, calculate speaker dependent Baum-Welch statistics and update speaker model $\mathcal{Y}_s$. \end{minipage} \\
\quad 3.2:
\begin{minipage}[t]{0.9\linewidth}
For each segment $m$ and speaker $s$, compute and update segment posteriors via eigenvoice scoring. \end{minipage} \\
\quad 3.3: Viterbi or forward-backward realignment with minimum duration constraint. \\
\quad 3.4 Repeat 3.1-3.3 until stopping criteria is met. \\
\hline
\end{tabular}
\end{center}
\end{table}

\section{Latent Class Model} \label{sec:lcm}
\begin{figure}[!ht]
	\centering
	\includegraphics [width=0.95\textwidth] {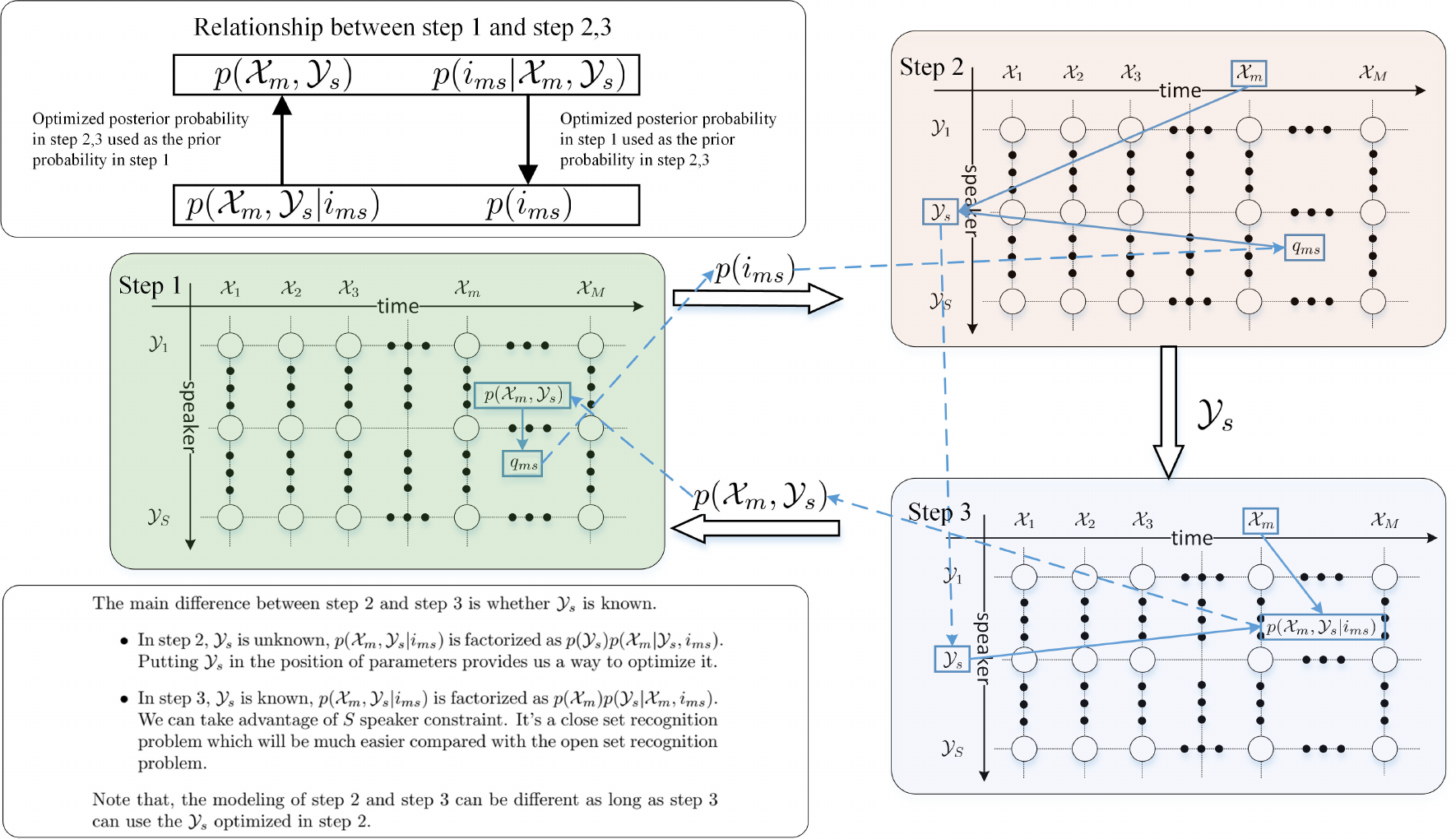}
	\caption{Diagram of LCM. The upper left text box illustrates the relationship between step 1 and step 2,3. The lower left text box explains the difference between step 2 and step 3.}
	\label{Figure02lcm_detail}.
\end{figure}

Suppose a sequence $\mathcal{X}$ is divided into $M$ segments, and $\mathcal{X}_m$ is the representation of segment $m$, $1\leq m \leq M$; $\mathcal{Y}_s$ is the representation of latent class $s$, $1\leq s \leq S$
Each segment belongs to one of $S$ independent latent classes.
This relationship is denoted by the latent class indicator matrix $I=\{i_{ms}\}$
\begin{equation} \label{equ:q_constrain}
  i_{ms} = \left\{
    \begin{array}{ll}
      1, & \text{if segment $m$ belongs to the latent class $s$} \\
      0, & \text{if segment $m$ does not belong to the latent class $s$} \\
    \end{array} \right.
\end{equation}
\noindent Our objective function is to maximizes the log-likelihood function with constraint that there are $S$ classes, as follows
\begin{equation}
  \begin{aligned}
    \arg_{Q, \mathcal{Y}} \max \log p(\mathcal{X}, \mathcal{Y}, I) & = \arg_{Q, \mathcal{Y}} \max  \sum_{m=1}^M \log \sum_{s=1}^S p(\mathcal{X}_m, \mathcal{Y}_s, i_{ms}) \\
    \text{s.t} & \quad \text{S  classes} \\
  \end{aligned}
\end{equation}
\noindent where $Q=\{q_{ms}\}$, $q_{ms}$ is the posterior probability which will be explained later. Intuitively, if $p(\mathcal{X}_m, \mathcal{Y}_s, i_{ms}) > p(\mathcal{X}_m, \mathcal{Y}_{s'}, i_{ms'}) , s' \neq s, 1 \leq s, s' \leq S $, we will draw a conclusion that segment $m$ belongs to class $s$. The above formula is intractable for the unknown $\mathcal{Y}$ and $I$.
We solve it through an iterative algorithm by introducing $Q$ as follows:
\begin{enumerate} \label{equ:obj}
	\item The objective function is factorized as
\begin{equation} \label{equ:obj_s1}
\begin{aligned}
    \sum_{m=1}^M \log \sum_{s=1}^S p(\mathcal{X}_m, \mathcal{Y}_s, i_{ms})
   & = \sum_{m=1}^M \log \sum_{s=1}^S p(\mathcal{X}_m, \mathcal{Y}_s) p(i_{ms}|\mathcal{X}_m, \mathcal{Y}_s) \\
   & = \sum_{m=1}^M \log \sum_{s=1}^S p(\mathcal{X}_m, \mathcal{Y}_s) q_{ms}
 \end{aligned}
 \end{equation}
In this step, $p(\mathcal{X}_m, \mathcal{Y}_s)$ is assumed to be known.
We use $q_{ms}$ denote $p(i_{ms}|\mathcal{X}_m, \mathcal{Y}_s)$ for simplicity.
Note that, $q_{ms} \geq 0$ and $\sum_{s=1}^S q_{ms} = 1$.
The (\ref{equ:obj})
is optimized by Jensen's inequality and Lagrange multiplier method. The updated $q_{ms}^{(u)}$ is
\begin{equation} \label{equ:update_q}
q_{ms}^{(u)} =\frac { q_{ms} p(\mathcal{X}_m, \mathcal{Y}_s )} { \sum_{s'=1}^S q_{ms'} p(\mathcal{X}_m, \mathcal{Y}_{s'})}
\end{equation}
The explanation for step 1 is that $q_{ms}$  is updated, given $p(\mathcal{X}_m, \mathcal{Y}_s)$ is known.
	
	\item The objective function is factorized as
	\begin{equation} \label{equ:obj_s2}
	\begin{aligned}
	\sum_{m=1}^M \log \sum_{s=1}^S p(\mathcal{X}_m, \mathcal{Y}_s, i_{ms})
	& = \sum_{m=1}^M \log \sum_{s=1}^S p(i_{ms}) p(\mathcal{X}_m, \mathcal{Y}_s|i_{ms}) \\
	& \approx \sum_{m=1}^M \log \sum_{s=1}^S q_{ms} p(\mathcal{Y}_s) p(\mathcal{X}_m |\mathcal{Y}_s, i_{ms})
	\end{aligned}
	\end{equation}
	There are two approximations used in this step. First, we use the posterior probability $q_{ms}$ in step 1 as the prior probability $p(i_{ms})$ in this step. Second, $p(\mathcal{Y}_s|i_{ms}) = p(\mathcal{Y}_s)$ is assumed.
	According to our understanding, $\mathcal{Y}_s$ is the speaker representation and $i_{ms}$ is the indicator between segment and speaker. Since $\mathcal{X}_m$ is not referenced,  $\mathcal{Y}_s$ and $i_{ms}$ are assumed to be independent of each other.
	A similar explanation is also given in Kenny's work, see (10) in \cite{Kenny2010}.
	The goal of this factorization is to put $\mathcal{Y}_s$ on the position of parameter, which provides a way to optimize it.
	And this step is to estimate $\mathcal{Y}_s$, given $p(i_{ms})$ is known.
	
	\item The objective function is factorized as
	\begin{equation} \label{equ:obj_s3}
	\begin{aligned}
	\sum_{m=1}^M \log \sum_{s=1}^S p(\mathcal{X}_m, \mathcal{Y}_s, i_{ms})
	& = \sum_{m=1}^M \log \sum_{s=1}^S p(i_{ms}) p(\mathcal{X}_m, \mathcal{Y}_s|i_{ms}) \\
	& \approx \sum_{m=1}^M \log \sum_{s=1}^S q_{ms} p(\mathcal{X}_m) p(\mathcal{Y}_s |\mathcal{X}_m, i_{ms})
	\end{aligned}
	\end{equation}
	
	There are also two approximations used in this step. First, we use the posterior probability $q_{ms}$ in step 1 as the prior probability $p(i_{ms})$ in this step.
	Second, $p(\mathcal{X}_m|i_{ms}) = p(\mathcal{X}_m)$ is assumed.
	According to our understanding, $\mathcal{X}_m$ is the segment representation and $i_{ms}$ is the indicator between segment $m$ and speaker $s$. Since $\mathcal{Y}_s$ is not referenced, $\mathcal{X}_m$ and $i_{ms}$ are assumed to be independent of each other.
	The explanation for step 3 is that $p(\mathcal{X}_m, \mathcal{Y}_s |i_{ms})$ is calculated, given $p(i_{ms})$ and $\mathcal{Y}_{s}$ are known. We compute the posterior probability $p(\mathcal{Y}_s |\mathcal{X}_m, i_{ms})$ rather than $p(\mathcal{X}_m |\mathcal{Y}_s, i_{ms})$
	to approximate $p(\mathcal{X}_m, \mathcal{Y}_s|i_{ms})$ with the goal that this factorization is to take advantages of $S$ speaker constraint.
	In next loop, $p(\mathcal{X}_m, \mathcal{Y}_s|i_{ms})$ is used as the approximation of $p(\mathcal{X}_m, \mathcal{Y}_s)$ and go to step 1, see Figure \ref{Figure02lcm_detail}.
\end{enumerate}

After a few iteration, the $q_{ms}$ is used to make the final binary decision.
We have several comments on the above iterations
\begin{itemize}
	\item Although the form of objective function ($\arg_{Q,\mathcal{Y}} \max \log p(\mathcal{X}, \mathcal{Y}, I)$) is the same in these three steps, the prior setting, factorized objective function and variables to be optimized are different, see Table \ref{tab:loop} and Figure \ref{Figure02lcm_detail}.
	This will also be further verified in the next section.

	\item The connection between step 1 and step 2,3 are $p(i_{ms})$ and $p(\mathcal{X}_m, \mathcal{Y}_s)$, see the upper left text box in Figure \ref{Figure02lcm_detail}. We use the posterior probability ($p(i_{ms}|\mathcal{X}_m, \mathcal{Y}_s)$ and $p(\mathcal{X}_m, \mathcal{Y}_s|i_{ms})$) in the previous step or loop as the prior probability ($p(i_{ms})$ and $p(\mathcal{X}_m, \mathcal{Y}_s)$) in the current step or loop.
	
	\item The main difference between step 2 and step 3 is whether $\mathcal{Y}_s$ is known, see the lower left text box in Figure \ref{Figure02lcm_detail}. The goal of step 2 is to make a more accurate estimation of speaker representation while the goal of step 3 is to compute $p(\mathcal{X}_m,\mathcal{Y}_s|i_{ms})$ in a more accurate way. The explicit functions in step 2 and step 3 can be different as long as $\mathcal{Y}_s$ is the same.
	
	\item A unified objective function or not? Not necessary. Of course, a unified objective function is more rigorous in theory, e.g VB \cite{Kenny2010}. In fact, we can use the above model to explain the VB in \cite{Kenny2010}. The (15), (19) and (14) in \cite{Kenny2010} are corresponding to step 1, 2 and 3, respectively \footnote{Note that, equal prior is assumed in (15) in \cite{Kenny2010}.}. However, the prior setting in each step is different, as stated in Table \ref{tab:loop}, we can take advantage of it to make a better estimation or computation. For example, we have two additional ways to improve $p(\mathcal{Y}_s, \mathcal{X}_m| i_{ms})$ in step 3, compared with the VB. First, the (14) in \cite{Kenny2010} is the eigenvoice scoring, given $\mathcal{X}_m$ and $\mathcal{Y}_s$ are known, which can be further improved by more effective scoring method, e.g. PLDA. Second, there are $S$ classes constraint, turning the open-set problem into the close-set problem.
		
	\item Whether the loop is converged? Not guaranteed. Since the estimation of $\mathcal{Y}_s$ and computation of $p(\mathcal{X}_m,\mathcal{Y}_s|i_{ms})$ are choices of designers, the loop will not converge for some poor implementation.
	But, if $ p^{u}(\mathcal{X}_m,\mathcal{Y}^{u}_{s^{*}}|i_{ms^{*}}=1) > p(\mathcal{X}_m,\mathcal{Y}_{s^{*}}|i_{ms^{*}}=1)$ (monotonically increase with upper bound) is satisfied, the loop will converge to a local or global optimal.
	The notation with star means that it's the ground truth. The $\mathcal{Y}$ with a superscript $u$ means the updated $\mathcal{Y}$ in step 2 and the $p$ with a superscript $u$ means another (or updated) similarity function in step 3.
	This also implies that we have two ways to optimize the objective function. One is to use a better $\mathcal{Y}$ (e.g. updated $\mathcal{Y}$ in step 2) and the other one is to choose a more effective similarity function.
		
	\item Whether the converged results conform to the diarization task? The Kullback-Leibler divergence between $Q$ and $I$ is
	$D_{\text{KL}}(I\|Q) = - \sum_{m=1}^M \log q_{ms}$.
	The minimization of KL divergence between $Q$ and $I$ is equal to the maximization of $\sum_{m=1}^M \log q_{ms}$. According to (\ref{equ:obj_s1}), $q_{ms}$ depends on $p(\mathcal{X}_m, \mathcal{Y}_{s})$.
	If $p(\mathcal{X}_m, \mathcal{Y}_{s^{*}}) > p(\mathcal{X}_m, \mathcal{Y}_{s'}), s^{*} \neq s'$ ($i_{ms^{*}}=1$ is the ground truth), the converged results will satisfy the diarization task.
	
	\item In addition to explicit unknown $Q$ and $\mathcal{Y}$, the unknown factors also include implicit functions, e.g. $p(\mathcal{X}_m,\mathcal{Y}_s| i_{ms})$ in step 2 and 3.
	These implicit functions are statistical models selected by designers in implementation.
	What we want to emphasize is that we can do optimization on its parameters for a already selected function, we can also do optimization by choosing more effective functions based on known setting, e.g. from eigenvoice to PLDA or SVM scoring.
	
\end{itemize}

\renewcommand{\arraystretch}{2}
\begin{table}[!ht]
	\caption{Settings for LCM in each step}
	\label{tab:loop}
	\centering
	\begin{threeparttable}
		\begin{tabular}{|l|c|c|c|}
			\hline
			Step & Prior setting & Factorized objective function & To be updated \\
			\hline
			1 & $ p(\mathcal{X}_m, \mathcal{Y}_s)$ & $\sum_{m=1}^M \log \sum_{s=1}^S p(\mathcal{X}_m, \mathcal{Y}_s) q_{ms} $ & $ q_{ms} $ \\
			\hline
			2 & $\mathcal{X}_m, q_{ms}$ & $\sum_{m=1}^M \log \sum_{s=1}^S q_{ms} p(\mathcal{X}_m| \mathcal{Y}_s, i_{ms}) p(\mathcal{Y}_s) $ & $\mathcal{Y}_s$ \\
			\hline
			3 & $\mathcal{X}_m,q_{ms},\mathcal{Y}_s$ & $\sum_{m=1}^M \log \sum_{s=1}^S q_{ms}  p(\mathcal{Y}_s|\mathcal{X}_m, i_{ms}) p(\mathcal{X}_m) $ & $ p(\mathcal{X}_m, \mathcal{Y}_s|i_{ms})$ \\
			\hline
		\end{tabular}
	\end{threeparttable}
\end{table}
\renewcommand{\arraystretch}{1.2}


\section{Implementation} \label{sec:realization}

If we regard speakers as latent classes, LCM will be a natural solution to a speaker diariazation task.
The implementation needs to solve three things further: specify the segment representation $\mathcal{X}_m$, specify the class representation $\mathcal{Y}_s$ and $p(\mathcal{X}_m, \mathcal{Y}_s)$ computation.
Depending on different considerations, they can incorporate different algorithms.
Given VB, LCM-Ivec-PLDA, LCM-Ivec-SVM as examples,

	\begin{enumerate}
		\item In VB, $\mathcal{X}_m$ is an acoustic feature. $\mathcal{Y}_s$ is specified as a speaker i-vector. $p(\mathcal{X}_m, \mathcal{Y}_s)$ is the eigenvoice scoring (Equation (14) in \cite{Kenny2010}).
		\item In LCM-Ivec-PLDA, $\mathcal{X}_m$ is specified as a segment i-vector.
		$\mathcal{Y}_s$ is specified as a speaker i-vector. $p(\mathcal{X}_m, \mathcal{Y}_s)$ is calculated by PLDA.
		\item In LCM-Ivec-SVM, $\mathcal{X}_m$ is specified as a segment i-vector. $\mathcal{Y}_s$ is specified as a SVM model trained on speaker i-vectors. $p(\mathcal{X}_m, \mathcal{Y}_s)$ is calculated by SVM .
	\end{enumerate}

\noindent Actually, $p(\mathcal{X}_m, \mathcal{Y}_s)$  can be regarded as a speaker verification task of short utterances, which will benefit from the large number of previous studies on speaker verification.

\begin{figure}[!ht]
\centering
\includegraphics [width=0.9\textwidth] {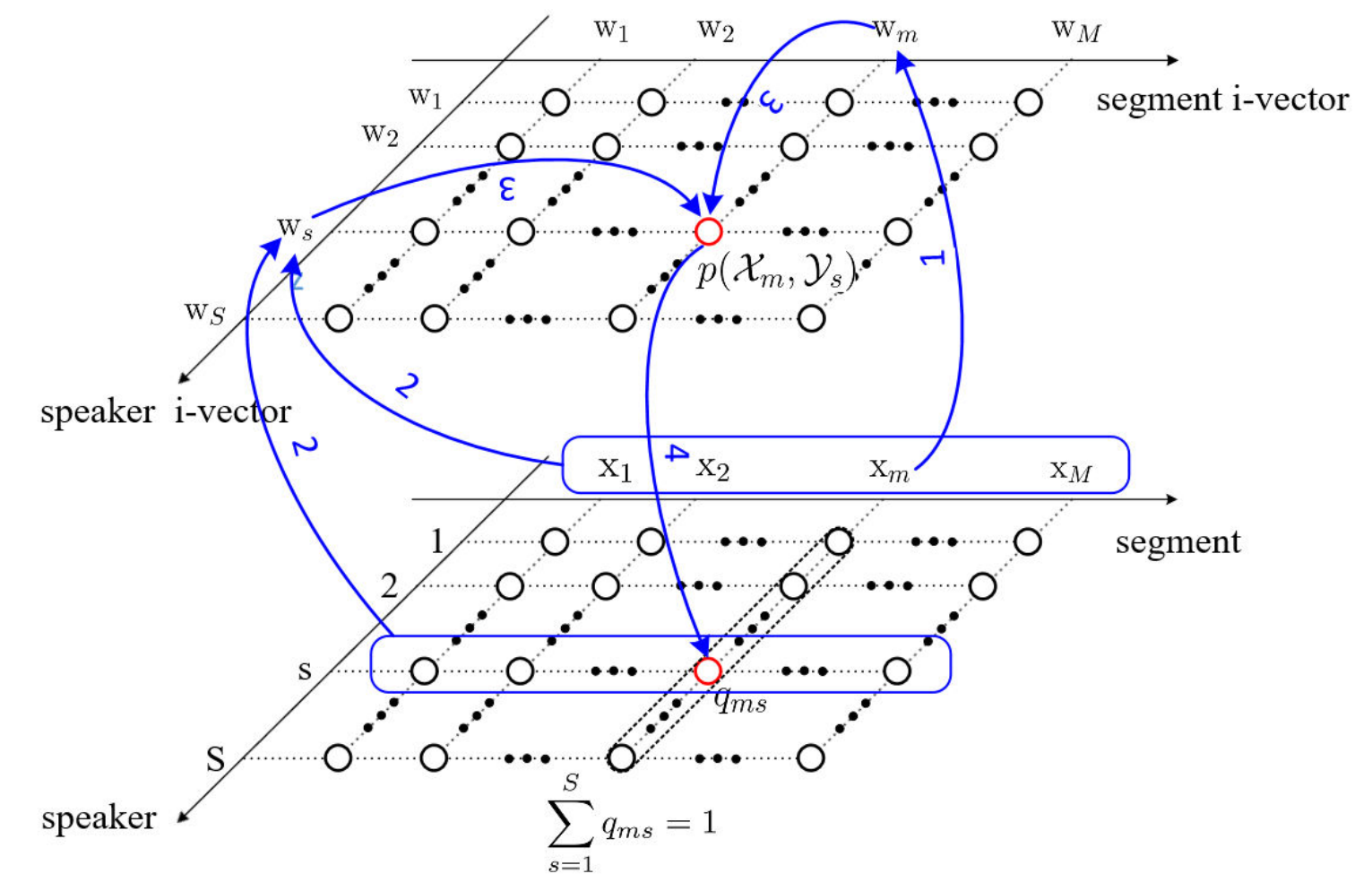}
\caption{Diagram of LCM speaker diarization. Step 1: Extract segment i-vector ${\rm{w}}_m$, Step 2: Extract speaker i-vector ${\rm{w}}_s$,  Step 3: Compute $p(\mathcal{X}_m, \mathcal{Y}_s)$, Step 4: Update $q_{ms}$.}
\label{Figure1fpm}
\end{figure}

The implementation of presented LCM-Ivec-PLDA speaker diarization is shown in Figure \ref{Figure1fpm}.
Different from the above section, $\mathcal{X}$ and $\mathcal{Y}$ are abstract representations of segment $m$ and speaker $s$.
In this section, they are specified to explicit expressions.
To avoid confusion, we use $\rm{x}$, $X$ and $\rm{w}$ to denote an acoustic feature vector, an acoustic feature matrix and an i-vector.
After front-end processing, the acoustic feature $X$ of a whole recording is evenly divided into $M$ segments, $X = \{{\rm{x}}_{1}, \cdots, {\rm{x}}_{M}\}$.
Based on the above notations, the iterative procedures of LCM-Ivec-PLDA is as follows (Figure \ref{Figure1fpm}):
\begin{enumerate}
\item segment i-vector ${\rm{w}}_m$ is extracted from ${\rm{x}}_{m}$ and its neighbors, which will be further explained in section \ref{sec:improvements}.
\item speaker i-vector ${\rm{w}}_s$ is estimated based on $Q = \{q_{ms}\}$ and $X =\{{\rm{x}}_{m}\}$.
\item $p(\mathcal{X}_m, \mathcal{Y}_s) = p(\rm{w}_m, \rm{w}_s)$ is computed through PLDA scoring.
\item Update $q_{ms}$ by $p(\mathcal{X}_m, \mathcal{Y}_s)$.
\end{enumerate}

This above process is repeated until the stopping criterion is met.
The step 1 is a standard i-vector extraction procedure \cite{Dehak2011} and step 4 is realized by (\ref{equ:update_q}).
So, we will put more attention on step 2 and 3 in the following subsections.

\subsection{Estimate speaker i-vector ${\rm{w}}_s$}
If $T$ denotes the total variability space, our objective function \cite{Kenny2010} is as follows
\begin{equation}\label{equ:ivec}
\begin{aligned}
&\arg_{\mathcal{Y}_s} \max \sum_{m=1}^M \log \sum_{s=1}^S q_{ms} p(\mathcal{X}_m, \mathcal{Y}_s) \\
=& \arg_{\rm{w}_s} \max \sum_{m=1}^M \log \sum_{s=1}^S q_{ms} p({\rm{x}}_m | {\rm{w}}_s) p({\rm{w}}_s) \\
=&\arg_{\rm{w}_s} \max \sum_{m=1}^M \log \sum_{s=1}^S q_{ms} \sum_{c=1}^{C} \omega_c \mathcal{N}({\rm{x}}_m | \mu_{\text{ubm},c} + T_{c} {\rm{w}}_s, \Sigma_{\text{ubm},c} )
\mathcal{N}({\rm{w}}_s|0, I_R)\\
 \end{aligned}
\end{equation}

\noindent where $C$ is the number of Gaussian mixture components.
$\mathcal{N}$ is a Gaussian distribution.
$\omega_c$, $\mu_{\text{ubm},c}$, and $\Sigma_{\text{ubm},c}$ are the weight, mean vector and covariance matrix of the $c$-th component of UBM, respectively.
$I_R$ is an identity matrix with rank $R$.
In contrast to speaker recognition in which the whole audio are assumed to be from one speaker, the segment $m$ belongs to speaker $s$ with a probability $q_{ms}$ in the case of speaker diarization.
We use Jensen's inequality \cite{jensen} again and obtain the lower bound as follows

\begin{equation}\label{equ:lowerLLF}
\begin{aligned}
\sum_{m=1}^{M} \sum_{s=1}^{S} q_{ms} \sum_{c=1}^C \gamma_{\text{ubm},mc} \log \mathcal{N}( {\rm{x}}_m | \mu_{\text{ubm},c} +  T_{c} {\rm{w}}_s, \Sigma_{\text{ubm},c} )\mathcal{N}({\rm{w}}_s|0, I_R)
\end{aligned}
\end{equation}

\noindent where
\begin{equation}
\begin{aligned}
\gamma_{\text{ubm},mc} = \frac{ \omega_{c} \mathcal{N}( {\rm{x}}_m | \mu_{\text{ubm},c}, \Sigma_{\text{ubm},c}) }{ \sum_{c'=1}^C \omega_{c'} \mathcal{N}( {\rm{x}}_m | \mu_{\text{ubm},c'}, \Sigma_{\text{ubm},c'}) }
\end{aligned}
\end{equation}
\noindent The above objective function is a quadratic optimization problem with the optimal solution
\begin{equation}\label{equ:esty}
{\rm{w}}_s = (I_R + T^t {N}_s \Sigma^{-1} T)^{-1} T^t \Sigma^{-1} {F}_s
\end{equation}
\noindent where ${N}_s$ and ${F}_s$ are concatenations of ${N}_{sc}$ and ${F}_{sc}$, respectively. $\Sigma$ is a diagonal matrix whose diagonal blocks are $\Sigma_{\text{ubm},m}$.
The ${N}_{sc}$, ${F}_{sc}$ are defined as follows
\begin{equation}
\begin{aligned}
{N}_{sc} &= \sum_{m=1}^{M} q_{ms} \gamma_{\text{ubm},mc} \\
{F}_{sc} &= \sum_{m=1}^{M} q_{ms} \gamma_{\text{ubm},mc} ({\rm{x}}_m -  \mu_{\text{ubm},c})\\
\end{aligned}
\end{equation}

In the above estimation, $T$ and $\Sigma$ are assumed to be known.
These can be estimated on a large auxiliary database in a traditional i-vector manner.

\subsection{Compute $p(\mathcal{X}_m, \mathcal{Y}_s)$}\label{sec:pms}
To compute $p(\mathcal{X}_m, \mathcal{Y}_s)$, we first extract segment i-vectors ${\rm{w}}_m$ from $\rm{x}_m$ and its neighbors, and evaluate the probability that ${\rm{w}}_m$ and ${\rm{w}}_s$ are from the same speaker.
We take advantages of PLDA and SVM to improve system performance, and propose LCM-Ivec-PLDA, LCM-Ivec-SVM and LCM-Ivec-Hybrid systems.

{\color{black}

\subsubsection{PLDA}

As each segment i-vector ${\rm{w}}_m$ and speaker i-vector ${\rm{w}}_s$ are known,
the task reduces to a short utterance speaker verification task at this stage.
We adopt a simplified PLDA \cite{simplified_plda} to model the distribution of i-vectors as follows:

\begin{equation} \label{equ:Ym}
{\rm{w}} = \mu_{I} + \Phi {\rm{y}} + \varepsilon
\end{equation}

\noindent where $\mu_{I}$ is the global mean of all preprocessed i-vectors, $\Phi$ is the speaker subspace, ${\rm{y}}$ is a latent speaker factor with a standard normal distribution, and residual term $\varepsilon \sim \mathcal{N} \left(0, \Sigma_\varepsilon \right)$. $\Sigma_\varepsilon$ is a full covariance matrix.
We adopt a two-covariance model and the PLDA scoring \cite{Br2010The,Prince2007} is
\begin{equation} \label{equ:PLDA}
s^{\text{PLDA}}_{ms} = \frac{p({\rm{w}}_m, {\rm{w}}_s \lvert i_{ms} = 1)}{p({\rm{w}}_m, {\rm{w}}_s \lvert i_{ms} \neq 1)},
\end{equation}
\noindent and the posterior probability with $S$ speaker constraint is
\begin{equation}
  \begin{aligned}
p(\mathcal{Y}_s | \mathcal{X}_m, i_{ms}) \propto \frac{(s^{\text{PLDA}}_{ms})^{\kappa}}{\sum_{s'=1}^S (s^{\text{PLDA}}_{ms'})^{\kappa}}
  \end{aligned}
\end{equation}
\noindent where $\kappa$ is a scale factor set by experiments ($\kappa = 1$ in the PLDA setting).
The explanation of $\kappa$ is similar to the $\kappa$ of (1) in \cite{Povey2008Boosted}.
As $p(\mathcal{X}_m)$ is the same for $S$ speakers and $p(\mathcal{Y}_s,\mathcal{X}_m| i_{ms}) = p(\mathcal{X}_m) p(\mathcal{Y}_s|\mathcal{X}_m, i_{ms}) $, the $p(\mathcal{X}_m)$ will be canceled in the following computation. The flow chart of LCM-Ivec-PLDA is shown in Figure \ref{Figure2lcm} without the flow path denoted as SVM.

\subsubsection{SVM}

Another discriminative option is using a support vector machine (SVM).
After the estimation of ${\rm{w}}_s$, we train SVM models for all speakers.
When training a SVM model ($\eta_s,b_s$) with a linear kernel for speaker $s$, ${\rm{w}}_s$ is regarded as a positive class and the other speakers $\omega_{s'}$($s'\neq s$) are regarded as negative classes. $\eta_s,b_s$ are linearly compressed weight and bias.

The SVM scoring is
\begin{equation}
s^{\text{SVM}}_{ms} = \eta_s {\rm{w}}_m + b_s
\end{equation}	
\noindent and the posterior probability with $S$ speaker constraint is
\begin{equation} \label{equ:SVM}
p(\mathcal{Y}_s | \mathcal{X}_m, i_{ms} ) \propto \frac{\exp(\kappa s^{\text{SVM}}_{ms})}{\exp(\kappa \sum_{s'=1}^{S} s^{\text{SVM}}_{ms'})}
\end{equation}
\noindent where $\kappa$ is a also scale factor ($\kappa = 10$ in the SVM setting).
As $p(\mathcal{X}_m)$ is the same for $S$ speakers and $p(\mathcal{Y}_s,\mathcal{X}_m| i_{ms}) = p(\mathcal{X}_m) p(\mathcal{Y}_s|\mathcal{X}_m, i_{ms}) $, the $p(\mathcal{X}_m)$ will be canceled in the following computation.
The flow chart of LCM-Ivec-SVM is shown in Figure \ref{Figure2lcm} without the flow path denoted as PLDA. }


\subsubsection{Hybrid}
\begin{figure}[!ht]
\centering
\includegraphics [width=0.7\textwidth]{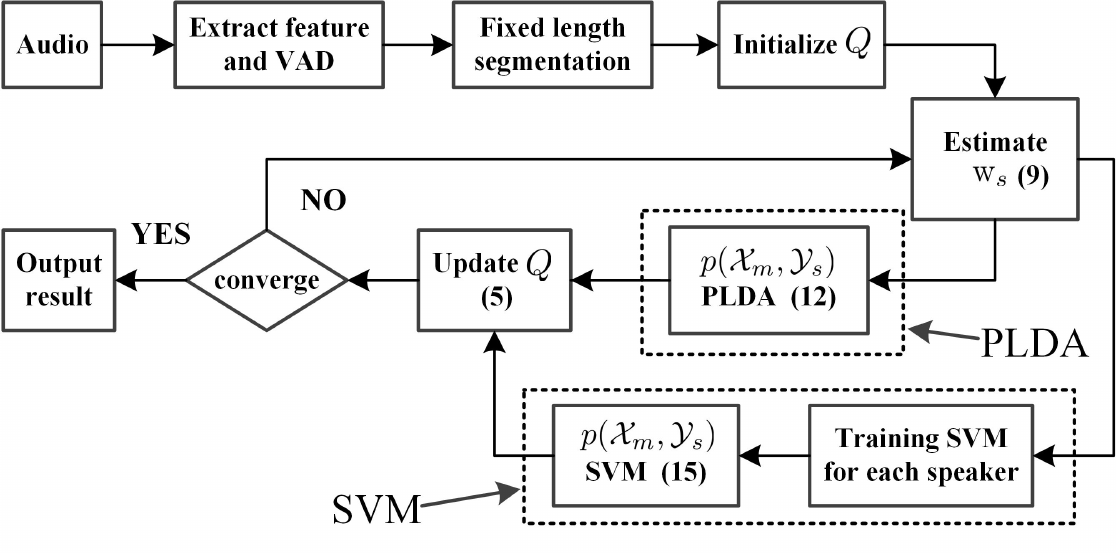}
\caption{Flow chart of LCM-Ivec-PLDA, LCM-Ivec-SVM and LCM-Ivec-Hybrid systems.}
\label{Figure2lcm}
\end{figure}

The calculation of $p(\mathcal{X}_m, \mathcal{Y}_s)$ is not explicitly specified in the LCM algorithm, which is just like the kernel function in SVM.
As long as the kernel matrix satisfies the Mercer criterion \cite{Mercer415}, different choices may make the algorithm more discriminative and more generalized.
In addition, multiple kernel learning is also possible by combining several kernels to boost the performance \cite{Bach:2004:MKL}.
In the LCM algorithm, as long as the probability $p(\mathcal{X}_m, \mathcal{Y}_s)$ satisfies the condition that the more likely both $\mathcal{X}_m$ and $\mathcal{Y}_s$ are from the same class $s$, the larger $p(\mathcal{X}_m, \mathcal{Y}_s)$ will be, we can take it and embrace more algorithms, e.g. the above mentioned PLDA and SVM.
We combine PLDA with SVM by iteration, see Figure \ref{Figure2lcm}.
This iteration takes advantages of both PLDA and SVM and is expected to reach a better performance.
This hybrid iterative system is denoted as LCM-Ivec-Hybrid system.

\begin{table}[!ht]
\begin{center}
\begin{tabular}{p{0.9\columnwidth}}
\hline
Algorithm 2: LCM-Ivec-PLDA, LCM-Ivec-SVM and LCM-Ivec-Hybrid \\
\hline
1: Voice activity detection and feature extraction \\
2: Segmentation \\
\quad 2.1: Split the audio into short segments equally, hence get $M$ segments.   \\
3: Clustering \\
\quad 3.1: Initialize $Q$ randomly \\
\quad 3.2: \begin{minipage}[t]{0.9\linewidth} Estimate speaker i-vector ${\rm{w}}_s$ (\ref{equ:esty}) based on $Q$ and $\rm{x}_m$ \end{minipage} \\
\quad 3.3: Extract each segment i-vector ${\rm{w}}_m$, see section \ref{sec:improvements} for more details. \\
\quad 3.4 (PLDA): \begin{minipage}[t]{0.8\linewidth} Calculate $p(\mathcal{X}_m, \mathcal{Y}_s)$ by PLDA (\ref{equ:PLDA}) for each segment and speaker. \end{minipage} \\
\quad 3.4 (SVM): \begin{minipage}[t]{0.8\linewidth} Train SVM for each speaker, and calculate $p(\mathcal{X}_m, \mathcal{Y}_s)$ by (\ref{equ:SVM}) for each segment and speaker. \end{minipage} \\
\quad 3.4 (Hybrid): do 3.4 (PLDA) and 3.4 (SVM) alternatively \\
\quad 3.5: Update $Q$ according to (\ref{equ:update_q}). \\
\quad 3.6: Repeat 3.2 - 3.5 until converge. \\
\hline
\end{tabular}
\end{center}
\end{table}

\section{Further Improvements} \label{sec:improvements}
\subsection{Neighbor Window}
In fixed length segmentation, each segment is usually very short to ensure its speaker homogeneity.
However, this shortness will lead to inaccuracy when extracting segment i-vectors and calculating $p(\mathcal{X}_m , \mathcal{Y}_s)$.
Intuitively, if a speaker $s$ appears at time $m$, the speaker will appear at a great probability in the vicinity of time $m$.
So its neighboring segments can be used to improve the accuracy of $p(\mathcal{X}_m , \mathcal{Y}_s)$.
We propose two methods of incorporating neighboring segment information.
At data level, we extract long term segmental i-vector $\mathcal{X}_m$ to use the neighbor information.
At score level, we build homogeneous Poisson point process model to calculate $p(\mathcal{X}_m , \mathcal{Y}_s)$.

\subsubsection{Data Level Window}
\begin{figure}[!ht]
	\centering
	\includegraphics [width=0.8\textwidth]{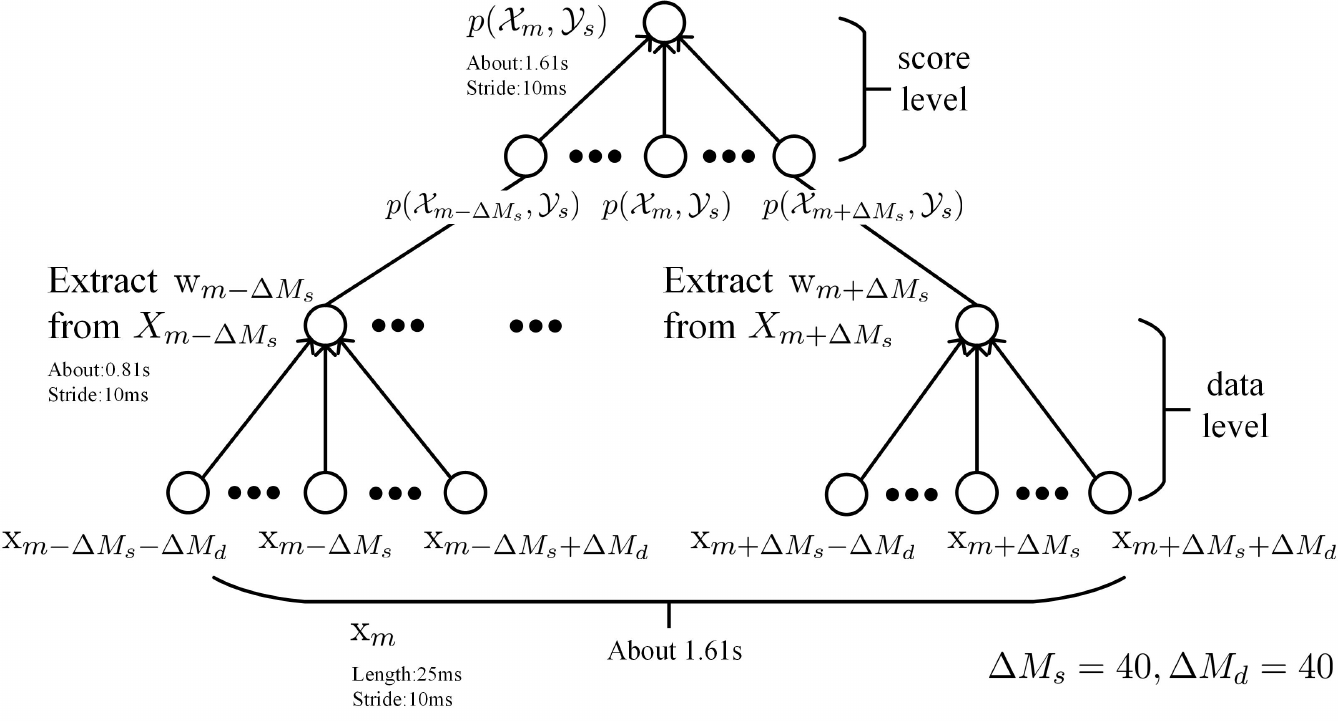}
	\caption{Data level and score level windows.}
	\label{Figure3nei}
\end{figure}

At the data level, we extract ${\rm{w}}_m$ using $\rm{x}_m$ and its neighbor data.
Let
\begin{equation}\label{equ:spk_nei_1}
X_m = \left ( {\rm{x}}_{m - \Delta M_{d}},\cdots,{\rm{x}}_m,\cdots, {\rm{x}}_{m + \Delta M_{d}} \right)
\end{equation}
\noindent where $ \Delta M_{d}$ is data level half window length, and $ \Delta M_{d} > 0$.
We use $X_m$ instead of ${\rm{x}}_m$ to extract i-vector ${\rm{w}}_m$ to represent segment $m$ as shown in the lower part of Figure \ref{Figure3nei}.
Since $X_m$ is long enough to ensure more robust estimates, system performance can be improved.
It should be noted that $X_m$ may contain more than one speaker, but this does not matter.
This is because the extracted ${\rm{w}}_m$ only represents the time $m$, not the time duration $\left(m-\Delta M_d, \cdots, m+\Delta M_d \right)$.
From another aspect, data level window can be seen as a sliding window with high overlapping to increase the segmentation resolution.

\subsubsection{Score Level Window}

\begin{figure}[!ht]
\centering
\includegraphics [width=0.6\textwidth] {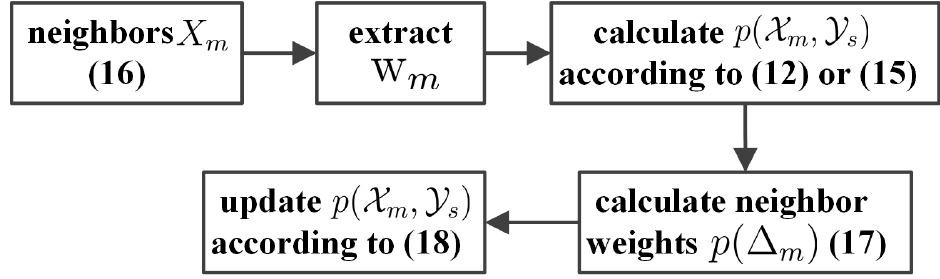}
\caption{Flow chart of adding neighbor window}
\label{Figure4neiflow}
\end{figure}

At the score level, we update $p(\mathcal{X}_m , \mathcal{Y}_s)$ with neighbor scores.
Given the condition that $m$-th segment belongs to speaker $s$, we consider the probability that $(m + \Delta m)$-th segment does not belong to speaker $s$.
If we define the appearance of a speaker change point as an event, the above process can be approximated as a homogeneous Poisson point process \cite{possion_point}.
Under this assumption, the probability that a speech segment from $m$ to $m + \Delta m$ belongs to the same speaker is equivalent to the probability that the speaker change point does not appear from $m$ to $m + \Delta m$, and can be expressed as:
\begin{equation} \label{equ:poisson}
  p (\Delta m ) = e^{ -\lambda \Delta m }, \Delta m \geq 0
\end{equation}
\noindent where $\lambda$ is the rate parameter.
It represents the average number of speaker change points in a unit time.
We consider the contribution of $p(\mathcal{X}_{m + \Delta m} , \mathcal{Y}_s)$  to $p(\mathcal{X}_m , \mathcal{Y}_s)$ by updating $p(\mathcal{X}_m , \mathcal{Y}_s)$ as follows,
\begin{equation}\label{equ:spk_nei_2}
  p(\mathcal{X}_m , \mathcal{Y}_s) \leftarrow \sum_{\Delta m = -\Delta M_s} ^{\Delta M_s} \left[ p (\Delta m ) p(\mathcal{X}_{m + \Delta m} , \mathcal{Y}_s)  \right]
\end{equation}
where $\Delta M_{s}$ is score level half window length, $\Delta M_{s} > 0$.
It should be noted that, $\Delta M_d$, $\Delta M_s$ and $\lambda$ are experiment parameters and will be examined in the next section.
As ${\rm{w}}_m$ is extracted from $X_m = ({\rm{x}}_{m - \Delta M_d}, \cdots, {\rm{x}}_{m + \Delta M_d})$,
in fact, the updated $p(\mathcal{X}_m , \mathcal{Y}_s)$ is related to $({\rm{x}}_{m - \Delta M_s - \Delta M_d}, \cdots, {\rm{x}}_{m + \Delta M_s + \Delta M_d})$,
as shown in Figure \ref{Figure3nei}.
The full process of incorporating two neighbor windows is shown in Figure \ref{Figure4neiflow}.


\subsection{HMM smoothing}
After several iterations, speaker diarization results can be obtained according to $q_{ms}$.
However, the sequence information is not considered in the LCM system, there might be a number of speaker change points in a short duration.
To address the frequent speaker change problem, a hidden Markov model (HMM) is applied to smooth the speaker change points.
The initial probability of HMM is $\pi_s = p(\mathcal{Y}_s)$.
The self-loop transition probability is $a_{ii}$ and the other transition probabilities are $a_{ij} = \frac{1-a_{ii}}{S-1}, i \neq j$.
Since the probability that a speaker transits to itself is much larger than that of changing to a new speaker,
the self-loop probability is set to be 0.98 in our work.
The emission probability is calculated based on PLDA (\ref{equ:PLDA}) or SVM (\ref{equ:SVM}).
With this HMM parameters, $q_{ms}$ can be smoothed using the forward-backward algorithm.

\subsection{AHC Initialization}
Although random initialization works well in most cases, LCM and VB systems tend to assign the segments to each speaker evenly in the case where a single speaker dominates the whole conversation, leading to poor results.
According to the comparative study \cite{CompareBottomUpTopDown2012}, we know that the bottom-up approach will capture comparatively purer models.
Therefore, we recommend an informative AHC initialization method, similar to our previous paper \cite{AHCprior2017}.
After using PLDA to compute the log likelihood ratio between two segment i-vectors \cite{Sell2014,Lan2016}, AHC is applied to perform clustering.
Using the AHC results, two prior calculation methods, hard prior and soft prior, are proposed \cite{AHCprior2017}.

\subsubsection{Hard Prior}
According to the AHC clustering results, if a segment $m$ is classified to a speaker $s$, we will assign $q_{ms}$ with a relatively larger value $q$.
The hard prior is as follows:
\begin{equation}\label{equ:hard prior}
q_{ms}= \mathcal{I} \left( {\mathcal{X}}_m \in s \right)q+ \mathcal{I} \left( {\mathcal{X}}_m \notin s \right)\frac{1-q}{S-1}
\end{equation}
\noindent where $\mathcal{I} \left(\cdot \right)$ is the indicator function.
$\mathcal{I} \left( {\mathcal{X}}_m \in s \right)$ means a segment $m$ is classified to speaker $s$.

\subsubsection{Soft Prior}b
For the soft prior, we first calculate the center of each estimated speaker $s$
\begin{equation}\label{equ:cluster center}
{\rm{\mu}}_{\rm{w}_s} = \frac{\sum_{m=1}^M \mathcal{I} \left({\rm{x}}_m \in s \right) {\rm{w}}_m} {\sum_{m=1}^M \mathcal{I} \left({\rm{x}}_m \in s \right)}
\end{equation}
The distance between ${\rm{w}}_m$ and ${\rm{\mu}}_{\rm{w}_s}$ is $d_{ms}=\lVert {\rm{w}}_m - {\rm{\mu}}_{\rm{w}_s} \rVert_2$.
According to the AHC clustering results, if a segment $m$ is classified to a speaker $s$, the prior probability for speaker $s$ at time $m$ is
\begin{equation}\label{equ:soft prior}
q_{ms}= \frac{1}{2}\left[ \frac{e^{- \left(\frac{d_{ms}}{d_{\text{max},s}} \right)^k}-e^{-1}}{1-e^{-1}}+ 1\right]
\end{equation}
\noindent where $d_{\text{max},s}=\max_{{\rm{x}}_m \in s} \left( d_{ms} \right)$, $k$ is a constant value.
This soft prior probability varies from 0.5 to 1, ensuring that if ${\rm{w}}_s$ is closer to ${\rm{\mu}}_{\rm{w}_s}$, $q_{ms}$ will be larger. For other speakers at time $m$, the prior probability is $(1-q_{ms})/(S-1)$.

\section{Related Work and Discussion} \label{sec:related}

\subsection{Core problem of speaker diarization}

Different from some mainstream approaches, we take a different view for the basic concept of speaker diarization.
Paper \cite{CompareBottomUpTopDown2012} summarized that the task of speaker diarization is formulated as solving the following objective function:
\begin{equation} \label{equ:obj_seg_clu}
\arg_{S,G} \max p(S,G|X)
\end{equation}
\noindent where $X$ is the observed data, $S$ and $G$ are speaker sequence and segmentation.
In our work, we formulate the speaker diarization problem as follows

\begin{equation} \label{equ:obj_ours}
\arg_{\mathcal{Y}, Q} \max p(\mathcal{X},\mathcal{Y},Q)
\end{equation}

\noindent where $\mathcal{X}$ be the observed data, $\mathcal{Y}$ and $Q$ are hidden speaker representation and latent class probability matrix.
Both objective functions can solve the problem of speaker diarization.
However, the objective function (\ref{equ:obj_seg_clu}) involves segmentation which introduces a premature hard decision that may degrade the system performance.
The objective function (\ref{equ:obj_ours}) has difficulty in solving speaker overlapping problem and depends on the accurate estimate of speaker number.

\subsection{Compared with VB}

 In VB, $\mathcal{Y}_s $ is a speaker i-vector and $p(\mathcal{X}_m , \mathcal{Y}_s)$ is the eigenvoice scoring (Equation (14) in [2]), a generative model.
In our paper, we replace eigenvoice scoring with PLDA or SVM scoring to compute $p(\mathcal{X}_m , \mathcal{Y}_s)$ which benefits from the discriminability of PLDA or SVM. Both VB and LCM-Ivec-PLDA/SVM are iterative processes, and there are two important steps:
\begin{description}
  \item[step 1] estimate $Q$  based on $\mathcal{X}$ and $\mathcal{Y}$.
  \item[step 2] estimate $\mathcal{Y}$ based on $\mathcal{X}$ and $Q$.
\end{description}
 The two algorithms are almost the same in the second step.
 However, in step 1, the calculation of $Q$ is more accurate by introducing the PLDA or SVM.
 In recent speaker recognition evaluations (e.g. NIST SREs), the Ivec-PLDA performed better than eigenvoice model (or joint factor analysis, JFA) \cite{Kenny2008}.
 The SVM is suitable for classification task with small samples.
 This is the reason why we introduce these two methods to LCM.
 Compared with VB, the main benefit of  LCM-Ivec-PLDA/SVM is that it takes advantages of PLDA or SVM to improve the accuracy of $p(\mathcal{X}_m , \mathcal{Y}_s)$.
 Besides, the $p(\mathcal{X}_m , \mathcal{Y}_s)$ is enhanced by its neighbors both at the data and score level.

\subsection{Compared with Ivec-PLDA-AHC}
The PLDA has many applications in speaker diarization.
Similar to GMM-BIC-AHC method, the Ivec-PLDA-AHC method has become popular in many research works.
This way of using i-vector and PLDA follows the idea of segmentation and clustering.
The role of PLDA is to evaluate the similarity of clusters divided by speaker change point, as done in paper \cite{ivector2015,Sell2014,Lan2016,PldaAhc2014,Zhu2016}.
Based on the PLDA similarity matrix, AHC is applied to the clustering task.
Although the performance is improved, it still has the premature hard decision problem.

\subsection{Compared with PLDA-VB}

In paper \cite{Bulut2015}, PLDA is combined with VB, and is similar to ours.
We believe that the probabilistic-based iterative framework, as depicted in the LCM, and not just the introduction of PLDA, is the key to solving the problem of speaker diarization.
Our subsequent experiments also prove that using SVM can achieve a similar performance.
The hybrid iteration inspired by the LCM can improve the performance further.
In addition, we also study the use of neighbor information, HMM smoothing and initialization method.

\section{Experiments} \label{sec:exp}

Experiments have been implemented on five databases: NIST RT09 SPKD SDM (RT09), our own speaker imbalanced \emph{TL} (\emph{TL}), LDC CALLHOME97 American English speech (CALLHOME97) \cite{ldc1997}, NIST SRE00 subset of the multilingual CALLHOME (CALLHOME00) and NIST SRE08 short2-summed (SRE08) databases to examine the performance of LCM.
Speaker error (SE) and diarization error rate (DER) are adopted as metrics to measure the system performance according to the RT09 evaluation plan \cite{RT09Eval} for RT09, \emph{TL}, CALLHOME97 and CALLHOME00 database.
Equal error rate (EER) and minimum detection cost function (MDCF08) are adopted as auxiliary metrics for SRE08 database.

\subsection{Common Configuration}
Perceptual linear predictive (PLP) features with 19 dimensions are extracted from the audio recordings using a 25 ms Hamming window and a 10 ms stride.
PLP and log-energy constitute a 20 dimensional basic feature.
This base feature along with its first derivatives are concatenated as our acoustic feature vector.
VAD is implemented using the frame log-energy and subband spectral entropy.
The UBM is composed of 512 diagonal Gaussian components.
The rank of the total variability matrix $T$ is 300.
For the PLDA, the rank of the subspace matrix is 150.
For segment neighbors, $\Delta M_d$, $\Delta M_s$ and $\lambda$ are 40, 40 and 0.05, respectively.

\subsection{Experiment Results with RT09}

The NIST RT09 SPKD database has 7 English meeting audio recordings and is about 3 hours in length.
The BeamformIt toolkit \cite{BeamformingIt} and Qualcomm-ICSI-OGI \cite{Adami02qualcomm-icsi-ogifeatures} front-end are adopted to realize acoustic beamforming and speech enhancement.
We use Switchboard-P1, RT05 and RT06 to train UBM, $T$ and PLDA parameters.
Three sets of experiments have been implemented to verify the performance of our proposed LCM systems, usage of neighbor window, and HMM smoothing on RT09 database, respectively.

\subsubsection{Comparison Among Different Methods}
In the first set of experiments, we study the performance of different systems on the RT09 database.
Table \ref{tab:Miss_and_FA} lists the miss (Miss) rate and false alarm (FA) speech rate of LCM-Ivec-Hybrid system.
It can be seen that the miss rate of the fifth recording reaches 20.0$\%$ percentage.
This recording has much overlapping speech which is not well handled by our proposed approach.

\begin{table}[!ht]
\caption{Miss and FA of LCM-Ivec-Hybrid system for RT09. Miss and FA are caused by VAD error and overlapping speech. They are very similar for all the three proposed systems, as the same VAD method is used.}
\label{tab:Miss_and_FA}
\centering
\begin{threeparttable}
\begin{tabular}{|l|c|c|}
\hline
       &Miss[\%]           &FA[\%]       \\
\hline
EDI\_20071128-1000  &3.64	&4.81		\\
\hline
EDI\_20071128-1500  &8.36	&6.68	 \\
\hline
IDI\_20090128-1600  &4.09	 &1.32	 \\
\hline
IDI\_20090129-1000  &5.91	&7.78	 \\
\hline
NIST\_20080201-1405  &20.01	&2.54 	 \\
\hline
NIST\_20080227-1501  &8.86	&1.26	 \\
\hline
NIST\_20080307-0955  &5.35	&2.49	  \\
\hline
average             &8.03	&3.84   \\
\hline
\end{tabular}
\end{threeparttable}
\end{table}

Results of GMM-BIC-AHC, VB and LCM-Ivec-PLDA/SVM/Hybrid systems are listed in Table \ref{tab:different method experiment}.
It can be seen that the performance of LCM systems is better than that of BIC system.
This can be ascribed to the usage of $q_{ms}$ for soft decisions instead of hard decisions.
The performance of LCM is also better than VB system.
This demonstrates that the introduction of a discriminative model is very effective.
VB is a method with an iterative optimization based on a generative model.
In contrast, LCM is a method with the computation of $p(\mathcal{X}_m, \mathcal{Y}_s)$ based on discriminative model,
which is in line with the basic requirements of the speaker diarization task and contributes to its performance improvement.
Compared with the classical VB system, the DER of LCM-Ivec-PLDA, LCM-Ivec-SVM, and LCM-Ivec-Hybrid have an average relative improvement of 23.5\%, 27.1\%, and 43.0\% on NIST RT09 database.
For some recordings, which already have good DERs with PLDA or SVM, the performance improvement of hybrid system is relatively small.
For others with poorer DERs, the improvement of the hybrid system is prominent.
We infer that the hybrid system may help to jump out of a local optimum achieved by a single algorithm.

\begin{table}[!ht]
\caption{Experiment results of different methods on RT09.}
\label{tab:different method experiment}
\centering
\begin{threeparttable}
\begin{tabular}{|l|c|c|c|c|c|c|c|}

\hline
\multicolumn{1}{|c|}{\multirow{2}{2.8cm}{DER[\%]}} &Speaker \#   &\multicolumn{1}{c|}{\multirow{2}{0.4cm}{BIC}} &\multicolumn{1}{c|}{\multirow{2}{0.4cm}{VB}}    & \multicolumn{3}{c|} {LCM-Ivec}  \\
\cline{5-7}
\multicolumn{1}{|c|}{}   &\multicolumn{1}{c|}{}  &\multicolumn{1}{c|}{}  &\multicolumn{1}{c|}{} &{PLDA}     &{SVM}     &{Hybrid}    \\
\hline
given speaker \#  &-  &Yes      &Yes	   &Yes		&Yes       &Yes\\
\hline
EDI\_20071128-1000  &4   &29.32      &10.67	   &9.89		&9.91       &9.83\\
\hline
EDI\_20071128-1500  &4  &35.61     &48.66		&19.68		&19.87   &17.40\\
\hline
IDI\_20090128-1600  &4   &29.12   	&11.15		&7.02		&7.14     &7.14 \\
\hline
IDI\_20090129-1000  &4  &37.27  	&35.85  	&31.99		&32.37    &21.82\\
\hline
NIST\_20080201-1405  &5  &61.54 	  &49.05	  &44.67	 &43.05   &38.53\\
\hline
NIST\_20080227-1501  &6  &40.32	     &39.97	     &24.76	   &25.66        & 13.96\\
\hline
NIST\_20080307-0955  &11  &46.62     	&23.50		&22.86		&16.44   & 16.00\\
\hline
average &-  &39.97     	&31.26		&22.98		&22.06     &17.81\\
\hline
\end{tabular}
\begin{tablenotes}
  \footnotesize
\item[1] The code for the BIC diarization system was downloaded from: https://github.com/gdebayan/Diarization$\_$BIC
\item[2] VB is the system described in P. Kenny's paper \cite{Kenny2010}. This system is partly realized by the python code downloaded from: http://speech.fit.vutbr.cz/software/vb-diarization-eigenvoice-and-hmm-priors.
\end{tablenotes}
\end{threeparttable}
\end{table}

We also compare our system performance with other research work in the literature.
Table \ref{tab:compared_with_others} lists the average performance of different methods on the RT09 database.
All of these systems except \cite{aIB2014} is under a SDM condition.
It can be seen that the Miss + FA of our method is relatively higher.
This is ascribed to the VAD error and overlapping speech.
Our method has the lowest SE and DER.

\begin{table}[!ht]
\caption{Compared with other work performance on RT09. Scoring overlapped speech is accounted in the error rates.}
\label{tab:compared_with_others}
\centering
\begin{threeparttable}
\begin{tabular}{|c|c|c|c|c|c|c|c|}
\hline
works        &approaches  &given speaker \#  &VAD[\%]   &Miss[\%]    &FA[\%]     & SE[\%]      &DER[\%]      \\
\hline
\cite{aIB2014}  &aIB    &No   &-   &11.6    &1.1		&14.3		&27.0 \\
\hline
\cite{IIR2010}  &GMM+BIC  &No  &2.7   &-     &-	&8.7		 &18.0        \\
\hline
\cite{ICSI2012}  &BottomUp   &No   &5.9  &-    &-		&-		&31.3 \\
\hline
\cite{EURECOM2010}  & TopDown   &No  &-    &-    &-		&-		&21.1 \\
\hline
\cite{CompareBottomUpTopDown2012}    &BottomUp+TopDown   &No   &9.0   &-    &-		&8.8	&17.8 \\
\hline
ours    &LCM  & Yes  &-     &8.0   &3.8 	&5.9	& 17.8  \\
\hline
\end{tabular}
\end{threeparttable}
\end{table}

\subsubsection{Effect of Different Neighbor Window}
\begin{figure}[!ht]
\centering
\includegraphics [width=0.5\textwidth] {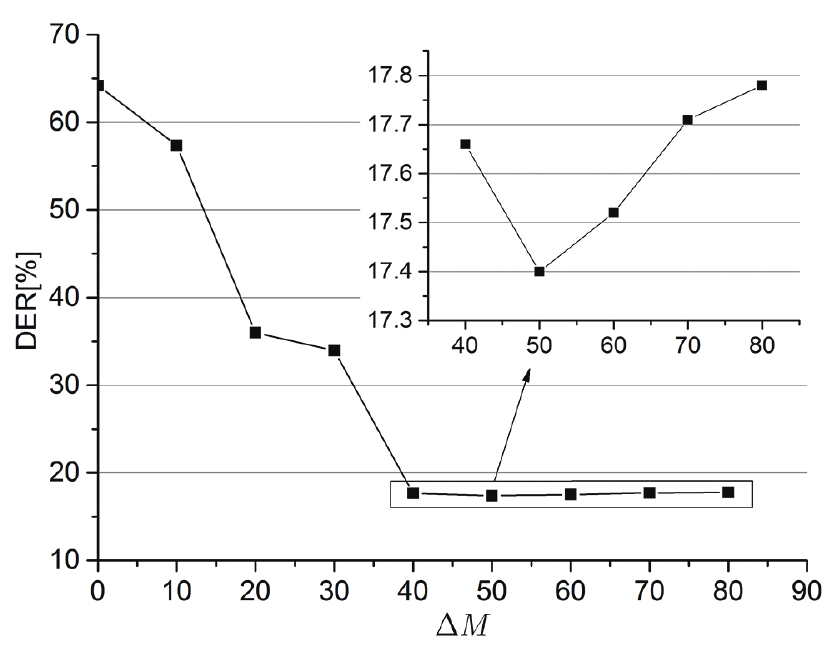}
\caption{DER varies with $\Delta M_d$ of data level window}
\label{Figure5dl}
\end{figure}

In the second set of experiments, we study the influence of different neighbor windows at both data level and score level.
For the data level window, Figure \ref{Figure5dl} shows the DER varies with $\Delta M_d$ of LCM-Ivec-Hybrid on the audio 'EDI\_20071128-1500'.
It can be seen that when $\Delta M_d = 0$, that is to say no data level window is added, the performance of the speaker diarization is poor.
As $\Delta M_d$ becomes larger, DER firstly decreases and then increases slightly.
This is because we can extract more speaker information from $\Delta M_d$ as it gets larger,
but if it grows too large, it begins to mix with other speaker's information.

\begin{figure}[!ht]
\centering
\includegraphics [width=0.7\textwidth] {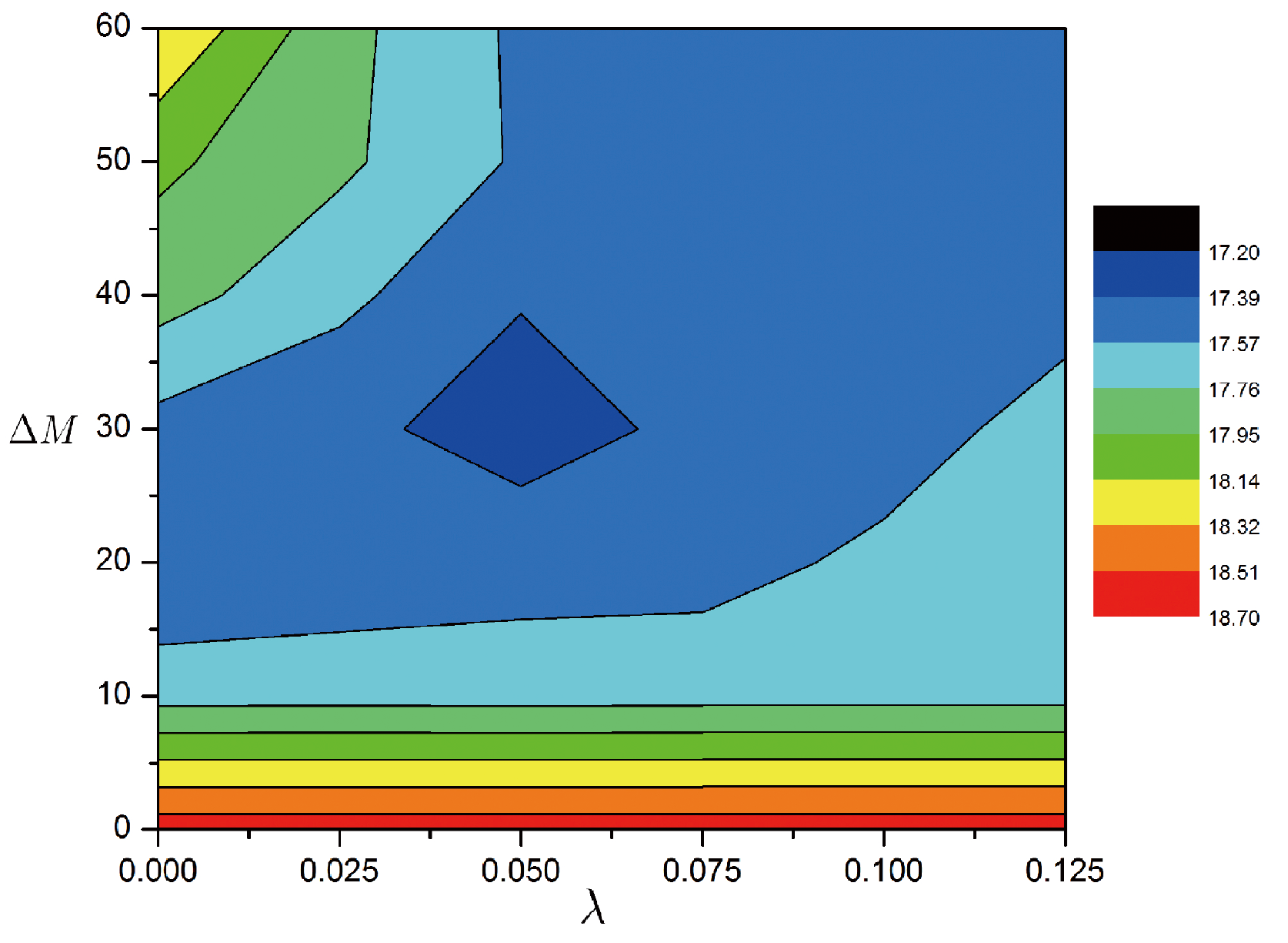}
\caption{DER varies with $\Delta M_s$ and $\lambda$ of score level window}
\label{Figure6sl}
\end{figure}

At the score level, the DER varied with $\Delta M_s$ and $\lambda$ is shown in Figure \ref{Figure6sl}.
We can see that,when $\lambda$ approaches to 0, the value of (\ref{equ:poisson}) approaches 1, and the Poisson window degrades to a rectangular window, DER also first decreases and then increases with $\Delta M_s$.
As $\lambda$ gets larger, the window becomes sharper, so DER is not so sensitive to a larger $\Delta M_s$.

Table \ref{tab:different neighbors experiment} shows the experimental results of the LCM system with or without neighbor windows on RT09.
All these systems are randomly initialized.
It can be seen that, from left to right, the performance of each system is gradually improved .
This demonstrates that taking segment neighbors into account improves the robustness and accuracy of $p(\mathcal{X}_m, \mathcal{Y}_s)$ both in LCM-Ivec-PLDA and LCM-Ivec-SVM systems,
thus enhancing the system performance.

\begin{table}[!ht]
\caption{Performance of LCM system with or without neighbor windows. The term 'no' means no neighbor window is added, while 'data' means adding only data level window, and 'data+score' means that both data and score level windows are added.}
\label{tab:different neighbors experiment}
\centering
\begin{threeparttable}
\begin{tabular}{|l|c|c|c|c|c|c|c|}
\hline
\multicolumn{1}{|l|}{DER[\%]}        & \multicolumn{3}{c} {LCM-Ivec-PLDA}   &\multicolumn{3}{|c|} {LCM-Ivec-SVM} \\
\hline
neighbor window           &no   &data    &data+score        &no   &data    &data+score \\
\hline
EDI\_20071128-1000   	&10.67	&10.66	&9.89	&10.72	&10.64	&9.91\\
\hline
EDI\_20071128-1500  	&45.14	&20.93	&19.68	&43.02	&20.77	&19.87 \\
\hline
IDI\_20090128-1600  	&11.38	&7.04	&7.02	&8.06	&7.61	&7.14 \\
\hline
IDI\_20090129-1000  	&34.00	&32.11	&31.99	&33.19	&32.24	&32.37 \\
\hline
NIST\_20080201-1405   &49.17	&49.17	&44.67	&44.43	&43.82	&43.05 \\
\hline
NIST\_20080227-1501  &58.49	 &47.11	&24.76	&27.01	&26.18	&25.66 \\
\hline
NIST\_20080307-0955  &24.91	&23.52	&22.86	&21.85	&20.44	&16.44 \\
\hline
\end{tabular}
\end{threeparttable}
\end{table}

\subsubsection{Effect of HMM Smoothing}

Table \ref{tab:HMM_smoothing} lists our third set of experiment results, from the LCM-Ivec-PLDA system with or without HMM smoothing.
It can be seen that, for the first six audio recordings, the SE and DER of the LCM-Ivec-PLDA system with HMM smoothing are better than that without HMM smoothing.
This can be ascribed to the HMM smoothing that makes the speaker changes less frequent.
For the seventh recording, the performance of LCM with HMM smoothing is not better than without HMM smoothing.
This is because the seventh recording has eleven speakers, and the speaker changes much more frequently than in the first six examples.
We guess that the HMM oversmooths the speaker change points, which means the loop probability is too large for this case.
In most cases, an HMM smoothing with proper parameters has positive effect.

\begin{table}[!ht]
\caption{Experiment result of LCM-Ivec-PLDA system with or without HMM smoothing}
\label{tab:HMM_smoothing}
\centering
\begin{threeparttable}
\begin{tabular}{|l|c|c|c|c|c|c|c|c|c|}
\hline
\multicolumn{1}{|c|}{\multirow{2}*{}}        & \multicolumn{2}{c} {SE[\%]}   &\multicolumn{2}{|c|} {DER[\%]} \\
\cline{2-5}
\multicolumn{1}{|c|}{}      &{noHMM}     &{HMM}     &{noHMM}      &{HMM}  \\
\hline
EDI\_20071128-1000  &1.5	   &1.4   &9.91 	&9.89  \\
\hline
EDI\_20071128-1500  &29.5	  &4.5   &44.68 	&19.68 \\
\hline
IDI\_20090128-1600  &12.1	  &1.7  &18.67	&7.02 \\
\hline
IDI\_20090129-1000  &13.4	  &12.1  &33.28	&31.99  \\
\hline
NIST\_20080201-1405  & 29.2	    &25.1  &48.72  	&44.67 \\
\hline
NIST\_20080227-1501  &14.8	    &13.7  &26.91	&24.76  \\
\hline
NIST\_20080307-0955  &10.1	    &14.6  &16.83	&22.86  \\
\hline
\end{tabular}
\end{threeparttable}
\end{table}

\subsection{Experiment Results with \emph{TL}}
The AHC initialization aims to solve of problem of speaker imbalance.
When there is one speaker dominating the whole conversation ($>80\%$ of the speech), VB and LCM will be sensitive to the initialization.
Random initialization results in poor performance.
But, if the conversation is not speaker imbalance, the initialization method has little influence on the performance.
All the experiments except this section are random initialized.

The AHC initialization experiment is carried out on our collected audio recordings \emph{TL}.
The training part of dataset \emph{TL} contains 57 speakers (30 female and 27 male).
The total duration is about 94 hours.
All of the recordings are natural conversations (Mandarin) recorded in a quiet office condition.
The evaluation part of \emph{TL} has 3 audio recordings (\emph{TL 7-9}).
These are also recorded in a quiet office, but there is one speaker who dominates the whole conversation ($>80\%$ of the speech).
Each recording has two speakers and is about 20 minutes.
In the AHC initialization, $q$ is set to be 0.7 in the hard prior setting and $k$ is 10 in the soft prior setting, unless explicitly stated.
Table \ref{tab:AHC_init} lists the SE and DER after AHC initialization before applying VB or LCM diarization.
The number of speakers is assumed to be known in advance.

\begin{table}[!ht]
\begin{center}
\caption{Experiment result of AHC initialization}
\label{tab:AHC_init}
\begin{tabular}{|c|c|c|}
\hline
AHC initial & SE[\%]  &DER[\%] \\
\hline
\emph{TL 7} &3.0 &5.9  \\
\hline
\emph{TL 8} &6.4 &11.4  \\
\hline
\emph{TL 9} &7.8 &9.5  \\
\hline
\end{tabular}
\end{center}
\end{table}

\begin{figure}[!ht]
\centering
\includegraphics [width=0.5\textwidth] {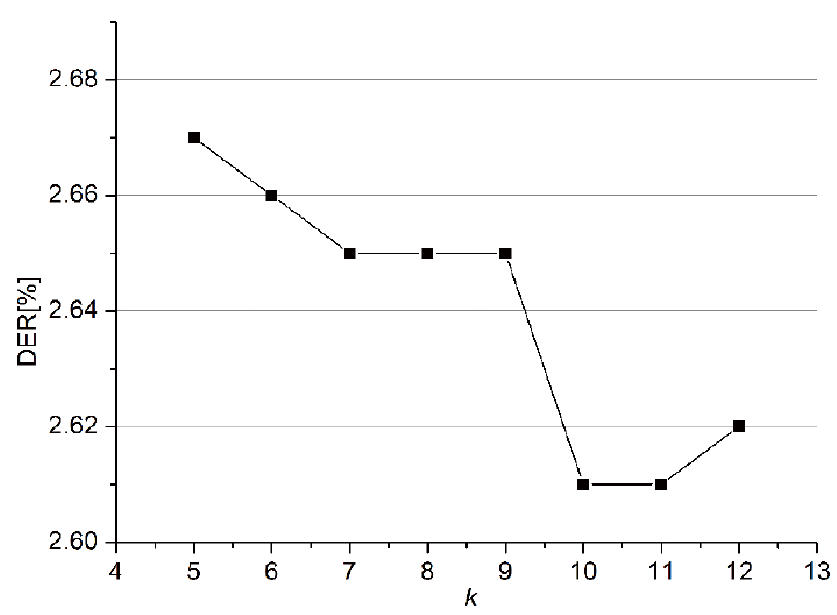}
\caption{DER varies with $k$ of soft prior (\ref{equ:soft prior})}
\label{Figure7vk}
\end{figure}

Figure. \ref{Figure7vk} shows the DER of 'TL 7' varies with $k$ of soft prior (\ref{equ:soft prior}).
According to the variation trend, we choose $k=10$ in our experiment.
From Table \ref{tab:Initialize Experiment}, we can see that random initialization gives poor results both in VB and LCM-Ivec-PLDA system in this case.
The proposed AHC hard and soft prior improves the system performance significantly.
The soft prior, which gives each segment an individual prior according to its distance to the estimated speaker centers, is more robust than the hard prior.
With the AHC initialization, the LCM-Ivec-PLDA and VB system both have significant improvement compared with their random prior systems.
The LCM-Ivec-PLDA system with hard/soft prior also surpasses the VB system with hard/soft prior with a relative improvement of 14.3$\%$/14.2$\%$.
able \ref{tab:AHC_init} and Table \ref{tab:Initialize Experiment} demonstrate that, although AHC initialization gets a not bad result, adding VB or LCM further improve the performance.

\begin{table}[!ht]
\begin{center}
\caption{Experiment result with random initialization and AHC initialization}
\label{tab:Initialize Experiment}
\begin{tabular}{|l|c|c|c|c|c|c|}
\hline
               & \multicolumn {3}{|c|}{SE[\%]} & \multicolumn {3}{|c|}{ DER[\%]}  \\
\hline
VB                      & random & hard prior & soft prior & random & hard prior & soft prior \\
\hline
\emph{TL 7}   & 36.9 & 1.7 & 1.9 & 40.1 &  4.9 & 5.2 \\
\hline
\emph{TL 8}   & 24.1 & 6.1 & 1.3 &  28.7 & 10.8 & 6.1 \\
\hline
\emph{TL 9}   & 30.6 & 6.6 & 1.1 & 32.4  &  8.4  & 2.9 \\
\hline
LCM-Ivec-PLDA            & random & hard prior & soft prior & random & hard prior & soft prior \\
\hline
\emph{TL 7}   & 38.8 & 8.5 & 0.6 & 42.0  & 10.5  & 2.6 \\
\hline
\emph{TL 8}   & 32.2 & 2.3 & 0.8 & 36.9  & 7.1  & 5.5 \\
\hline
\emph{TL 9}   & 44.7 & 6.2 & 1.1 & 46.5  & 8.0   & 2.9 \\
\hline
\end{tabular}
\end{center}
\end{table}

\subsection{Experiment Results with CALLHOME97}
The LDC CALLHOME97 American English speech database (CALLHOME97) consists of 120 conversations.
Each conversation is about 30 minutes and includes about 10 minute transcription.
Only the transcribed parts are used.
There are 109, 9 and 2 conversations containing 2, 3 and 4 speakers, respectively.
We follow the practice of \cite{Lapidot2017} and \cite{Zajic2016}, conversations with 2 speakers are examined.
We use Switchboard P1-3/Cell and SRE04-06 to train UBM, $T$ and PLDA parameters.

\begin{figure}[!ht]
\centering
\includegraphics [width=0.95 \textwidth] {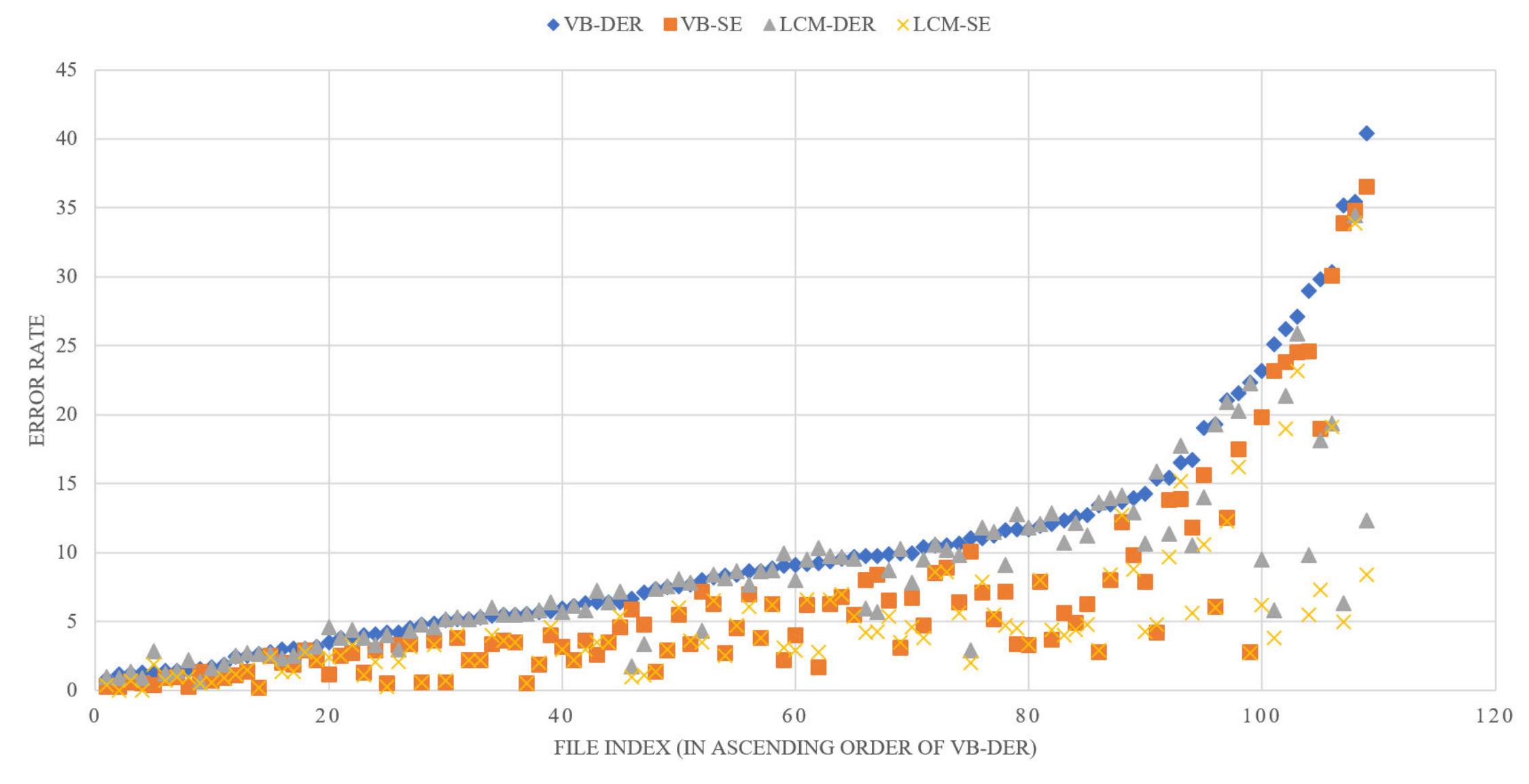}
\caption{DER and SE of VB and LCM-Ivec-Hybrid on CALLHOME97 database}
\label{Figure8ch}
\end{figure}

Scatter chart \ref{Figure8ch} enumerates VB-DER (blue diamond), VB-SE (orange square), LCM-Ivec-Hybrid-DER (LCM-DER, grey triangle) and LCM-Ivec-Hybrid-SE (LCM-SE, yellow cross) in the ascending order of VB-DER.
Both LCM-DER and LCM-SE are lower than VB-DER and VB-SE in summary, see also Table \ref{tab:compared_with_others_callhome}.

We find an interesting thing.
In the low region of DER ($<6\%$), the performance of VB and LCM systems is similar.
In the middle-to-high region of DER ($>6\%$), LCM is not better than VB for all test conversations, but it has a significant performance improvement for a considerable number of conversations, see the distribution of blue diamonds and grey triangles in Figure \ref{Figure8ch}.
The same situation is also reflected in Table \ref{tab:different method experiment}.
We believe that the VB is trapped in a local optimum for these segments.
By contrast, the LCM avoids this situation by incorporating with different methods.
In addition, the standard deviation of DER and SE of the LCM is smaller (Table \ref{tab:compared_with_others_callhome}), indicating that the performance of the LCM system is more stable.

Table \ref{tab:compared_with_others_callhome} compares the results.
It can be seen that compared with the VB system, the LCM-Ivec-Hybrid system has a relatively improvement of $26.6\%$ and $17.3\%$ in SE and DER, respectively.
Compared with other listed methods, the LCM-Ivec-Hybrid system also performs best on the CALLHOME97 database.
Diarization systems based on i-vector, VB or LCM are trained in advance and perform well in fixed conditions. While diarization systems based on HDM have little prior training, it can perform better if test conditions vary frequently.

\begin{table}[!ht]
\caption{Comparison with other works on CALLHOME97 database.}
\label{tab:compared_with_others_callhome}
\centering
\begin{threeparttable}
\begin{tabular}{|l|c|c|c|}
\hline
works                                    & method                                         & SE[\%]        & DER[\%]      \\
\hline
\cite{Lapidot2017} 	             & hidden distortion models (HDM)	 &   -           & 12.71       \\
\hline
\cite{Zajic2016}  	                 & GMM-Ivec                                   & 	 -          & 9.8   \\
\hline
\cite{Kenny2010} + ours        & VB + windows                             & 6.58 $\pm$ 7.59 & 10.08 $\pm$ 8.09 \\
\hline
 ours                                      & LCM-Ivec-Hybrid                        & 4.84 $\pm$ 4.95 & 8.33  $\pm$ 5.83 \\
\hline
\end{tabular}
\end{threeparttable}
\end{table}

\subsection{Experiment Results with CALLHOME00}
The CALLHOME00, a subtask of NIST SRE00, is a multilingual telephone database and consists of 500 recordings.
Each recording is about $2 \sim 5$ minutes in duration, containing $2 \sim 7$ speakers.
We use oracle speech activity marks and speaker numbers.
Similar to \cite{Castaldo2008Stream,Shum2013,Senoussaoui2013A,Sell2014,Povey_ASRU2011}, overlapping error is not accounted.
So, the DER is identical to the SE in this section.
We use Switchboard P1-3/Cell and SRE04-06 to train UBM, $T$ and PLDA parameters.

From Table \ref{tab:compared_with_others_callhome00}, we may draw a conclusion that our proposed methods are optimal.
However, it's not fair for \cite{Castaldo2008Stream, Senoussaoui2013A,Sell2014,Povey_ASRU2011}.
Paper \cite{Castaldo2008Stream, Povey_ASRU2011} don't use the oracle VAD and paper \cite{Castaldo2008Stream, Senoussaoui2013A,Sell2014} don't use the oracle speaker number.
And both two factors have a great influence on the system performance.
These results can only be used as an auxiliary reference.
Paper \cite{Shum2013} has the same setting with our work and the proposed  LCM-Ivec-Hybrid is slightly better.
Based on the results of above three sections, we guess that our proposed system is more suitable for long speech, for the reason that $\mathcal{Y}_s$ can be estimated more accurately from the long speech.

\begin{table}[!ht]
\caption{Result (in DER$[\%]$) on CALLHOME00 database.}
\label{tab:compared_with_others_callhome00}
\centering
\begin{threeparttable}
\begin{tabular}{|l|c|c|c|c|c|c|c|}
\hline
speaker \#                                         & 2 (303) & 3 (136) & 4 (43) & 5 (10) & 6 (6) & 7 (2) & Average \\
\hline
Table 2 in \cite{Castaldo2008Stream} & 8.7 & 15.7 & 15.1 & 20.2 & 25.5 & 29.8 & 11.67 \\
\hline
Figure 5 in \cite{Shum2013} *            & 5.0 &  12.5 & 17.7 & 20.5 & 21.5 &  33.1   &  8.75 \\
\hline
Table 5 in \cite{Senoussaoui2013A}   & 7.5 &  11.8 & 14.9 & 22.8 & 25.9 &  26.9   & 9.91 \\
\hline
 \cite{Sell2014}                                  & - & - & - & - & - & - & 13.7 \\
\hline
Kaldi \cite{Povey_ASRU2011}          & - & - & - & - & - & - & 8.69 \\
\hline
VB+windows                                     & 6.68 & 14.51 & 18.68 &  26.78 & 24.88 & 25.35 & 10.53 \\
\hline
 LCM-Ivec-Hybrid                              & 4.26 & 13.12 & 17.96 & 25.74 & 24.70 & 25.35 & 8.60 \\
\hline
\end{tabular}
\end{threeparttable}
\begin{tablenotes}
\footnotesize
\item[1] () denotes the number of recordings.
\item[2] * reflects that these numbers are measured from figures.
\end{tablenotes}
\end{table}

\subsection{Experiment Results with SRE08}
The NIST SRE08 short2-summed channel telephone data consists of 1788 models and 2215 test segments.
Each segment is about 5 minutes in duration (about 200 hours in total).
We find that there is no official speaker diarization key for the summed data.
Thus, neither DER or SE is adopted for this set of experiments.
The paper \cite{Kenny2010} reports that "We see that there is some correlation between EER and DER, but this is relatively weak".
So, we measure the effect of diarization through EER and MDCF08 in an indirect way.
On one hand, we use the NIST official trials (short2-summed, short2-summed-eng).
On the other hand, we follow the practice of \cite{Vaquero2013Quality} and make extended trials (ext-short2-summed, ext-short2-summed-eng).

We use Switchboard P1-3/Cell and SRE04-06 to train UBM, $T$ and PLDA parameters.
Here, our speaker verification system is a traditional GMM-Ivec-PLDA system.
The extracted 39 dimension PLP feature has 13 dimension static feature, $\Delta$ and $\Delta\Delta$.
A diagonal GMM with 2048 components is gender-independent.
The rank of the total variability matrix $T$ is 600.
For the PLDA, the rank of the subspace matrix is 150 \cite{simplified_plda}.

To begin with, we give some experimental results on the NIST SRE08 core tasks, i.e. short2-short3-telephone (short2-short3) and short2-short3-telephone-English trials (short2-shor3-eng), to verify the performance of above speaker verification system, see Table \ref{tab:compared_with_others_summed}.
Compared with the classical paper \cite{Dehak2011}, our results are normal.
Subsequently, we present results of the same speaker verification system on the NIST SRE08 short2-summed condition.
Without the front diariazation, the EER and MDCF08 are as high as $16.94\%$ and $0.686$.
Whether it is a VB + windows or LCM-Ivec-Hybrid, speaker diarization can significantly improve system performance.
Comparing case 5,9,14,17 with case 6,10,15,18 in Table \ref{tab:compared_with_others_summed}, we think that the performance improvement of LCM over VB is mainly due to the better diarization of LCM.

According to our literature research, there are few documents that report EER and MDCF08 on the short2-summed condition.
We list state-of-the-art diariation-verification systems developed by the LPT \cite{Dalmasso2009Loquendo,Castaldo2011Loquendo} in 2008 in Table \ref{tab:compared_with_others_summed}.
Paper \cite{Kenny2010} also presents the related EER in its Figure 4.
Compared with them, our system works better.
Part of the reason is the advance of speaker verification system, and the other part is the effectiveness of our proposed methods.

Paper \cite{Vaquero2013Quality} gives results on the extended trials which is more convincing in our opinion.
On the ext-short2-summed trials, although our EER ($4.99\%$) is worse than their report ($4.39\%$), but our MDCF08 ($0.201$) is better than their report ($0.209$).
Besides, paper \cite{Vaquero2013Quality} is a fusion system but our work is a single system.

\begin{table}[!ht]
  \caption{Results on NIST SRE08 summed channel telephone data}
  \label{tab:compared_with_others_summed}
  \centering
  \begin{threeparttable}
    \begin{tabular}{|l|l|c|c|c|}
      \hline
      Case & Trials (Ivec-PLDA)      & Diarization    & EER[\%]      &MDCF08      \\
      \hline
      \hline
      1 &       short2-short3           & - 	                                             & 4.47     & 0.245      \\
      \hline
      2 &       short2-summed           & -            	                                 & 16.94    & 0.686       \\
      \hline
      3 &       short2-summed           & Figure 4 in \cite{Kenny2010}     & 9.0         & -              \\
      \hline
      4 &       short2-summed           & LPT  \cite{Dalmasso2009Loquendo,Castaldo2011Loquendo}    & -            & 0.493       \\
      \hline
      5 &       short2-summed           & VB + windows                          & 9.64     & 0.410       \\
      \hline
      6 &       short2-summed           & LCM-Ivec-Hybrid                     & 8.71     & 0.374       \\
      \hline
      \hline
      7 &       ext-short2-summed       & - 	                                         & 10.77    & 0.438       \\
      \hline
      8 &       ext-short2-summed       & Table 2 in \cite{Vaquero2013Quality}                & 4.39        & 0.209       \\
      \hline
      9 &       ext-short2-summed       & VB + windows                        & 5.48     & 0.228       \\
      \hline
      10 &       ext-short2-summed       & LCM-Ivec-Hybrid                   & 4.99     & 0.201       \\
      \hline
      \hline
      11 &       short2-short3-eng           & - 	                                         & 1.76     & 0.0895       \\
      \hline
      12 &       short2-summed-eng           & - 	                                         & 14.25    & 0.504       \\
       \hline
      13 &       short2-summed-eng           & LPT  \cite{Dalmasso2009Loquendo,Castaldo2011Loquendo}   & -            & 0.282       \\
      \hline
      14 &       short2-summed-eng           & VB + windows                       & 6.33     & 0.236       \\
      \hline
      15 &       short2-summed-eng           & LCM-Ivec-Hybrid                  & 5.62     & 0.245       \\
      \hline
      \hline
      16 &       ext-short2-summed-eng       & - 	                                        & 10.00    & 0.400       \\
      \hline
      17 &       ext-short2-summed-eng       & VB + windows                     & 4.13    &  0.154       \\
      \hline
      18 &       ext-short2-summed-eng       & LCM-Ivec-Hybrid                 & 3.48     & 0.133       \\
      \hline
    \end{tabular}
  \end{threeparttable}
\end{table}

\section{Conclusion} \label{sec:con}
In this paper, we have applied a latent class model (LCM) to the task of speaker diarization.
LCM provides a framework that allows multiple models to compute the probability $p(\mathcal{X}_m, \mathcal{Y}_s)$.
Based on this algorithm, additional LCM-Ivec-PLDA, LCM-Ivec-SVM and LCM-Ivec-Hybrid systems are introduced.
These approaches significantly outperform traditional systems.

There are five main reasons for this improvement:
1) Introducing a latent class model to speaker diarization and using discriminative models in the computation of $p(\mathcal{X}_m, \mathcal{Y}_s)$ which enhances the system's ability at distinguishing speakers.
2) Incorporating temporal context through neighbor windows, which increases speaker information extracted from each short segment.
This incorporation is used both at the data level, taking $\mathcal{X}_m$ and its neighbors to constitute $X_m$ when extracting $\mathcal{Y}_m$, and at the score level, considering the contribution of neighbors when calculating $p(\mathcal{X}_m, \mathcal{Y}_s)$.
3) Performing HMM smoothing, which takes the audio sequence information into consideration.
4) AHC initialization is also a crucial factor when the conversation is dominated by a single speaker.
5) The hybrid schema can avoid the algorithm falling into local optimum in some cases.

Finally, our proposed system has the best overall performance on NIST RT09, CALLHOME97, CALLHOME00 and SRE08 short2-summed database.

\begin{backmatter}

\section*{Ethics approval and consent to participate}
Not applicable.

\section*{Consent for publication}
Not applicable.

\section*{Availability of data and material}
The CALLHOME00, NIST RT05/RT06/RT09 and NIST SRE08 database were released by NIST during evaluation.
The above mentioned data and LDC CALLHOME97 can also be obtained by LDC (https://www.ldc.upenn.edu/).
For the TL database, please contact author for data requests (heliang@mail.tsinghua.edu.cn).

\section*{Funding}
The work was supported by the National Natural Science Foundation of China under Grant No. 61403224.

\section*{Author's contributions}
The contributions of each author are as follows:
\begin{itemize}
  \item Liang He proposed the LCM methods, score level windowing and the AHC initialization for speaker diarization.
  He realized these systems in C++ code (Aurora3 Project) and wrote the paper.
  \item Xianhong Chen did a lot of literature research, wrote the section of mainstream approaches and algorithms and studied the data level windowing and HMM smoothing.
  \item Can Xu did experiments on the database based on the code provided by He Liang.
  He also recorded and analyzed the results.
  \item Yi Liu provided the comparative systems.
  \item Jia Liu gave some advices.
  \item Michael T Johnson checked the whole article and polished English language usage.
\end{itemize}

\section*{Acknowledgements}
We would like to thank the editor and anonymous reviewers for their careful work and thoughtful suggestions that have helped improve this paper substantially.

\section*{Competing interest}
The authors declare that they have no competing interests.

\bibliographystyle{bmc-mathphys}
\bibliography{20180719_lcm}

\setcounter{figure}{0}
\setcounter{table}{0}

\end{backmatter}
\end{document}